\documentclass[12pt]{emulateapj}
\usepackage{lscape}

\submitted{To Appear in Astronomical Journal.}

\newcommand{\etal}{{\it et al.}}

\newcommand{\hicont}{$\langle {\rm HI} \rangle$}

\shorttitle{H$\alpha$ Scale Lengths}

\begin{document}
\title
{A Comparison of H$\alpha$ and Stellar Scale Lengths in Virgo and Field Spirals}

\author{Rebecca A. Koopmann}
\affil{Union College Department of Physics and Astronomy, Schenectady, NY 12308}
\email{koopmanr@union.edu}

\author{Martha P. Haynes}
\affil{Center for Radiophysics and Space Research and National Astronomy
and Ionosphere Center\footnote{
The National Astronomy and Ionosphere Center is operated by Cornell University
under a cooperative agreement with the National Science Foundation.}, 
530 Space Sciences Building, Cornell University, Ithaca, NY 14853}
\email{haynes@astro.cornell.edu}

\author{Barbara Catinella}
\affil{National Astronomy and Ionosphere Center, 
Arecibo Observatory, HC3 Box 53995, Arecibo, PR 00612}
\email{bcatinel@naic.edu}

\begin{abstract}

The scale lengths of the old stars and ionized gas distributions are
compared for similar samples of Virgo Cluster members and field spiral
galaxies via H$\alpha$ and broad R--band surface photometry. While the
R--band and H$\alpha$ scale lengths are, on average, comparable for
the combined sample, we find significant differences between the field
and cluster samples.  While the H$\alpha$ scale lengths of the field
galaxies are a factor of 1.14 $\pm$ 0.07 longer on average than their
R--band scale lengths, the H$\alpha$ scale lengths of Virgo Cluster
members are on average 20\% smaller than their R--band scale lengths.
Furthermore, in Virgo, the scale length ratios are correlated with the
size of the star-forming disk: galaxies with smaller overall H$\alpha$
extents also show steeper radial falloff of star formation
activity. At the same time, we find no strong trends in scale length
ratio as a function of other galaxy properties including galaxy
luminosity, inclination, morphological type, central R--band light
concentration, or bar type.  Our results for H$\alpha$ emission are
similar to other results for dust emission, suggesting that H$\alpha$
and dust have similar distributions.  The environmental dependence of
the H$\alpha$ scale length places additional constraints on the
evolutionary process(es) that cause gas depletion and a suppression of
the star formation rate in clusters of galaxies.

\end{abstract}


\section{Introduction}

A quantitative comparison of the distribution of current star formation 
sites to that of the older stellar population in normal spiral galaxies can
place important constraints on the laws of star formation in spiral disks and 
the  star formation history of the universe.

Stellar disks of spiral galaxies follow an exponential distribution
in their surface brightness but what of the star formation? 
Schmidt (1959) suggested that the radial trend of star formation 
could be described by a power law of the gas density.
According to such ``Schmidt laws'', 
the distribution of star formation should follow the distribution of gas. 
Kennicutt (1989) and Martin \& Kennicutt (2001) found that the
gas surface density, normalized by disk area, is correlated with the star
formation rate, as long as the gas surface density exceeds a critical
value set by gravitational instability. The radial distributions
of molecular (e.g., Nishiyama, Nakai, \& Kuno 2001; Regan \etal ~2001; 
Young \etal ~1995) and atomic (e.g., Warmels 1988; Cayatte \etal ~1994;
Thomas \etal ~2004) gas have been shown for many galaxies to be approximately
exponential over much of a galaxy's disk. Molecular and atomic gas 
scale lengths have been shown to correlate with stellar scale lengths 
(Regan \etal ~2001), with dust scale lengths (Thomas \etal ~2004) and with
each other (Nishiyama \etal ~2001), although the sample sizes are small
and the scatter large. 
The main issues discussed in this paper are (a) whether
the radial behavior of the sites of on-going massive star formation 
is similar to that exhibited by the older stellar disk, under the
assumption that both are fairly approximated by exponential functions,
and (b) whether their distributions depend on the galaxy environment.

A common tracer of massive star formation arising from the youngest
stellar population (age $< 10$ Myr) is the H$\alpha$ emission line,
which can be used to estimate the total star formation rate 
in relatively dust-free regions by making 
standard assumptions about the initial mass function (Kennicutt 1983).
It is relatively easy to acquire H$\alpha$ observations for
large samples of galaxies, and therefore the comparison of H$\alpha$ and 
broadband scale lengths can be used to probe the linkage between
the young, massive stars and the older, more evolved stellar population.

Previous studies have found that the radial distribution of massive star 
formation as traced by H$\alpha$ is indeed exponential over large areas 
of the disk, but they have often presented conflicting results on the relative 
values of H$\alpha$ and stellar exponential scale lengths.
Hodge \& Kennicutt (1983) found for 14 spirals 
that H$\alpha$ scale lengths, measured from HII region 
distributions, are comparable to broadband scale lengths. Athanassoula,
Garc\'ia G\'omez, \& Bosma (1993), also using HII region distributions,
found that H$\alpha$ scale lengths are on average 10\% (20\%)
\it shorter \rm than those of stellar scale lengths, as measured in the 
red (blue), for their sample of 64 spiral galaxies.
Ryder and Dopita (1994; hereafter RD94) fit scale lengths to H$\alpha$ 
surface brightness profiles derived from CCD imaging for a sample of 
34 nearby spirals, finding H$\alpha$ scale lengths 75\% and 60\% \it longer \rm
on average than those in broadband I and V, respectively. 
The observation of longer H$\alpha$ scale lengths is consistent with
a number of studies documenting color gradients in spiral disks, such
that spirals become bluer with increasing radius (e.g., de Jong 1996b)
due to dust and/or age/metallicity effects (de Jong 1996b;
Peletier \etal ~1994; de Grijs 1998; Cunow 1998, 2001).
However, it is surprising that the difference in relative H$\alpha$ scale
lengths as measured by different authors should be so large. In addition,
broadband scale length ratios tend to be significantly smaller than the 
H$\alpha$:I ones found by RD94; for example, literature
values for the B:I ratio range from 1.12 to 1.36
for galaxies of low to high inclination, respectively (de Jong 1996b; 
de Grijs 1998).

In the hopes of resolving this conflict, we compare in this paper
H$\alpha$ and R--band scale lengths for a sample of 103 spiral galaxies.
The sample of galaxies is summarized in Section \ref{galsamp}. 
Section \ref{sfit} presents a description of the
fitting procedures used to extract both R--band and H$\alpha$ scale lengths
from imaging data. The comparison of the scale lengths 
is discussed in Section~\ref{relative}, while Section~\ref{compare}
investigates the relationship
between H$\alpha$ scale lengths and other galaxy properties, including
the size of the star-forming disk, HI content, and the distributions
of molecular gas and dust. Section~\ref{summ} provides a summary and
discussion of our results.

\section{Sample of Galaxies}
\label{galsamp}

For this study, we have made use of existing imaging and spectroscopic data for
two sets of galaxies:
a sample of 52 Virgo cluster spirals and a complementary set of 51 similar 
field galaxies. 
Broadband R and H$\alpha$ imaging data for the cluster sample
were extracted from Koopmann 
\etal ~(2001), who selected bright ($B_T^0$ $<$ 13.5) members
from the Virgo Cluster Catalog (Binggeli \etal 1985).
The field sample is made up of spirals outside of clusters 
(based on studies such
as Tully 1987) and was compiled from several sources of broad-- and 
narrow--band imaging data available to us. It includes
24 spirals from the isolated galaxy sample of Koopmann \& Kenney 
(2005),  14 from the studies of van Zee \etal ~(1998) and Jore (1997;
see also Haynes
\etal ~2000), and 13 from the Palomar/Las Campanas Atlas of Nearby 
Galaxies (Koopmann, R. A. \etal, ~in prep.).
Details of the image reduction procedures are given in the cited papers. 
We note that  no correction was made for contamination by the 
2 [NII] $\lambda\lambda$6548, 6584 \AA lines
that also lie within the bandpasses of the H$\alpha$ filters. Thus
H$\alpha$ in this paper should be read as H$\alpha$+[NII] (see also
Koopmann \& Kenney 2005).
For the Koopmann \etal ~(2001) and Koopmann \& Kenney (2005) datasets,
surface photometry was obtained from the R--band and continuum-subtracted 
H$\alpha$ images using fixed center, inclination, and
position angle elliptical apertures, following techniques discussed
in Koopmann \etal ~(2001). The H$\alpha$ images from 
the van Zee \etal ~and Jore datasets were constructed from 
ON-- and OFF--line narrow band images which were then analyzed
following procedures identical to those used for the Koopmann ones.

Table~\ref{sampcomp} provides a comparison of the mean distances, luminosities,
axial ratios, Hubble types, and bar fractions of galaxies in the 
two environments. 
The Virgo objects are placed at the mean cluster
distance of 16.6 Mpc, while distances to the field objects were calculated
adopting the local flow model of Tonry \etal ~(2000) using a program kindly
provided by K.L. Masters. Based on this model, the field galaxies have
a mean distance of 21.4 $\pm$ 1.4 Mpc.
Because distance errors calculated from
such a model carry a large error for individual objects located in the local
volume, most of the analysis performed in this work relies on
distance-independent relative quantities.

Face-on total blue magnitudes, B$_T^0$, were extracted from
the \it Third Reference Catalog \rm (de Vaucouleurs \etal ~1991, 
hereafter the RC3) and $M_B$ values were calculated using the derived 
distances. The luminosity ranges of the two samples are similar, but the 
field sample is brighter in the mean by 0.7 mag. 
The inclination range and mean of the two
samples are similar, as can be seen from the values of the mean
axial ratios.

Hubble types are extracted from two sources: 
(i) the RC3 and (ii) the \it Revised Shapley Ames Catalog \rm 
(Sandage \& Tammann 1987, hereafter RSA) 
supplemented by  the Binggeli \etal ~(1985) and Sandage \& Bedke (1994) 
catalogs. We will collectively refer to types extracted from the latter 3 
references as `RSA' types.
Both samples span a range across the Hubble
spiral sequence but have comparable mean values of the RC3 type
index $<$T$>$ and a similarly calculated RSA type index.
The bar fractions, derived from the RC3 and given as a ratio
of percent of sample in Table~\ref{sampcomp}, are similar for the two sets,
although the field sample contains a relatively higher number of strongly
barred galaxies.

\begin{deluxetable}{lcc}
\tabletypesize{\scriptsize}
\tablecaption{Sample Comparison\label{sampcomp}}
\tablewidth{0pt}
\tablehead{
\colhead{Property}& 
\colhead{Field}&
\colhead{Virgo}
}
\startdata
Distance (Mpc) & $21.4 \pm 1.4$ & 16.6 \\
B$_T^0$ & $11.1 \pm 0.2$&$11.6 \pm 0.1$\\
$M_B$ & $-20.3 \pm 0.1$ & $-19.6 \pm 0.1$\\
Axial Ratio & $0.62 \pm 0.02$ & $0.65 \pm 0.03$\\
RSA $<$T$>$ & $3.9 \pm 0.2$& $4.1 \pm 0.2$\\
RC3 $<$T$>$ & $4.0 \pm 0.3$ & $3.8 \pm 0.4$\\ 
SA:SAB:SB (\%) & 27:37:35 & 36:36:27 \\
\enddata
\end{deluxetable}

\section{Derivation of R--band and H$\alpha$ Scale Lengths}\label{sfit}

In this section, we discuss
how the scale lengths were derived, particularly in the case of the
H$\alpha$ surface distributions which, by their patchy nature, are more
complicated to fit.

\subsection{R--band Scale Lengths}\label{Rscale}
R--band disk scale lengths, $r_R$, were determined using the bulge-disk 
decomposition utility
within the IRAF routine \it nfit1d. \rm  
Since we were most interested in disk scale lengths, no attempt was
made to determine best-fit bulge parameters. This approach is based
on the observation that the disk scale length is the most stable parameter
derived in a bulge-disk decomposition (de Jong 1996a).
The exponential disk fit was made over as large a range in galaxy
radius as possible while avoiding the bulge region.
We estimate the uncertainty in R--band scale lengths derived in this way to
be 10-15\% (see Koopmann \etal ~2001).

While more stable than bulge parameters, disk scale lengths are known
to be dependent on the fitting process. Knapen \& 
van der Kruit (1991) find that scale lengths are extremely sensitive to the 
radial range fit and that literature values for the same galaxy differ
on average by 23\%, with a scatter of 20\%. 
M\"ollenhoff (2004) provides an example of the scatter
for the spiral NGC~4321. Scale lengths are also dependent on extinction
within galaxies (Giovanelli \etal ~1994). 
The galaxies in both our Virgo and field samples are of moderate inclination
(mostly $40^\circ < i < 70^\circ$), and thus we make no corrections for 
extinction in this paper, 
but compare our results as a function of inclination to check for 
extinction effects.

When we compare our R--band scale lengths to those of Grosb$\o$l (1985) 
for 21 field and 24 Virgo galaxies in common, 
we find that our R--band scale lengths are the same on average 
as those of Grosb$\o$l (1985)
with a scatter of 25\%. (Two very discrepant Virgo galaxies, NGC 4321 and
NGC 4698, were removed from the sample because Grosb$\o$l's much smaller
values were obviously measured from a steeper inner portion of the radial
profile.)

\subsection{H$\alpha$ Scale Lengths}\label{hascale}
Because H$\alpha$ emission traces the youngest stellar population, at an age
of less than 10 Myr, the HII region distribution is expected to 
strongly reflect the spiral arm pattern
in grand-design spirals and to be quite
patchy in more flocculent systems. In many objects, the H$\alpha$ emission
peaks in ring structures corresponding to spiral resonances and is often
not traceable away from the localized star formation sites. Because of the 
irregular
nature of the HII distribution, even the azimuthally averaged
surface brightness profiles of the H$\alpha$ emission 
are characteristically less smooth than those traced by 
the significantly older stellar population responsible for the R--band
isophotes. Hence, deriving the scale length from the H$\alpha$ images
is substantially more difficult than the similar task for the relatively
smooth stellar ones. 

We start from the basic assumption that the star-forming disk has an 
underlying exponential distribution, albeit locally modified by the patchy
nature of the massive star formation process.
Ideally the H$\alpha$ scale length, $r_{\alpha}$, should be fit over as much 
of a galaxy's
star-forming disk as possible. Yet a more limited radial range must be 
adopted to insure a consistent fit for all galaxies in the sample and to 
avoid systematic errors from inner regions, which are subject to 
uncertainties resulting from active nuclei, bulge continuum subtraction, 
dust, and clearing of gas by the presence of a bar, and outer regions, which
are measured at lower sensitivity. We therefore use the scale lengths derived
from the broadband images, $r_R$, to set the inner and outer limits for the
H$\alpha$ fits. The radial fitting range was based on R--band scale 
lengths rather than isophotal radii because photometrically
calibrated observations were not available for a large portion of the 
field sample. We note that the broadband scale length  
is correlated with isophotal measures of galaxy extent, such that the 
characteristic isophotal radii in various bands (i.e., r$_{25}$ in B, 
r$_{24}$ in R,
r$_{23.5}$ in I) typically lie between 3-4 scale lengths 
(e.g., Giovanelli \etal ~1995; Koopmann \etal ~2001).
For galaxies with absolute photometry, fits were made
within radial ranges defined by the isophotal radii, e.g., 0.3-0.9r$_{24}$, 
and compared to the results obtained using ranges defined by the R--band scale
length. There were no significant differences in the results.

To avoid contamination from the bulge, the inner regions 
of the galaxy at $r < 1r_R$ are excluded from the fits. 
As for the choice of the outer boundary for the fitting range, we
adopted three different options. \\
(1) {\em $r=3r_R$} -- 
Most of the field galaxies exhibit H$\alpha$ emission of sufficiently 
high S/N out to $\sim3r_R$, so we use this as a first outer boundary.
However, a large number of the galaxies in our sample, 
especially those in the Virgo
sample, have star-forming disks that are truncated 
near or within 3$r_R$
(e.g., Koopmann \etal ~2001; Dale \etal ~2001; Koopmann \& Kenney 2004a).
There are also several galaxies, mostly Virgo,
that have resolved HII regions extending to or beyond 3$r_R$, but because
of a small number of HII regions, the 
azimuthally-averaged H$\alpha$ surface brightness 
profile approaches or is below the estimated sky uncertainty over the
outer part of the 1-3$r_R$ range.
We elected to include these latter galaxies in the sample fit between
1-3$r_R$. To test the effect of including these galaxies, we also 
performed fits over the radial interval specified next. \\
(2) {\em $r=r_s$} -- In this case we fit all the H$\alpha$ profile
points between $r_R$ and $r_s$, the radius at which the uncertainty is
equal to the uncertainty in the sky.
This fit also had the advantages of 
(i) allowing a wider radial fitting range in the case of many galaxies, 
(ii) including several more Virgo galaxies in which the H$\alpha$
emission is truncated at a radius between 2 and 3 $r_R$, and 
(iii) allowing a comparison to previous measurements of H$\alpha$ scale length 
(e.g., RD94). \\
(3) {\em $r=2r_R$} -- 
Finally, we also made a fit to the more limited range of 1-2$r_R$ 
for all the galaxies with H$\alpha$ emission over this region, in order
to test for differences between the inner and outer disk scale lengths.

H$\alpha$ profiles tend to be ``bumpy'' due to the presence of
bright HII regions, rings of star formation, and regions swept of gas by a 
bar. These features can systematically affect fits made over a limited 
radial range. Thus we also fit scale lengths to profiles that had been 
radially
smoothed. Smoothing was done within IDL using a multiple boxcar smoothing 
technique (\it smooth2\rm)  from the Johns Hopkins University library of 
IDL routines.
A disadvantage of the smoothing process is that information from regions
outside the 1-3$r_R$ range, i.e., bulge, bar, and outer (often truncated)
disk, is included.  For galaxies with approximately 
exponential profiles, the unsmoothed and smoothed fits are similar.
For those that deviate from exponential profiles,
the unsmoothed and smoothed fits can
be significantly different. The fitting range and degree of smoothing then
become critical to the scale length estimate. 

Figure~\ref{examfits} illustrates these points for 
two example galaxies in our field sample, NGC 3359 (upper) and NGC
2712 (lower). The R--band and H$\alpha$ (continuum-subtracted) images of the
galaxies are shown on the left, next to the derived surface
brightness profiles (right panels, solid lines; the H$\alpha$ profile
is displaced by an arbitrary offset for display purpose). As mentioned above,
the patchy H$\alpha$ emission yields a profile that is significantly
less smooth than the corresponding R--band one. For both galaxies, the
R--band profile outside the innermost 1-2 kpc is well described by
an exponential function (dashed). The situation is more
complicated for the H$\alpha$ profiles, which are shown in both
smoothed (dotted) and unsmoothed (solid) version. Also shown are the
exponential fits in the radial intervals 1-3$r_R$ (solid) and $r_R-r_s$
(dashed) from which two of the three types of H$\alpha$ scale lengths discussed
above are derived. In the case of NGC 2712 (as well as in other,
similar figures presented in this paper), the $r_R-r_s$ fitting range
does not include the outermost points of the profile, since their
uncertainties are above the sky noise fluctuations.
For objects like NGC 3359, with a fairly exponential H$\alpha$
profile, the derived scale length is similar no matter the smoothing or
fitting range. 
The H$\alpha$ radial profile of NGC 2712, on the other
hand, has a shape reminiscent of a Freeman Type II profile, caused by
the ring of star formation and bar region cleared of H$\alpha$ emission.
The derived scale lengths depend strongly on the range chosen for the fit
and the degree of smoothing. We judge that it is better to exclude such
galaxies from our sample than to guess at a best fit. 
Galaxies like NGC 2712, in which the fit either produced a positive
slope or which otherwise obviously deviated from exponential,
were therefore eliminated. The fits to all galaxies were examined by eye 
and best-fit scale lengths and errors assigned. In most cases, the value
of the best fit was close to the average of the smoothed and unsmoothed fit
and the error encompassed the values from the smoothed and unsmoothed
fits. In some cases, either the smoothed or unsmoothed fit was 
deemed to be more reliable
and was therefore weighted more heavily in the final result. In the
comparisons of H$\alpha$ and R--band scale lengths, galaxies with errors
larger than 20\% were rejected, thus
effectively eliminating the objects with the least 
exponential profiles.

\section{Comparison of Stellar and H$\alpha$ Scale Lengths}\label{relative}

\begin{figure*}
\centering
\includegraphics[scale=0.55]{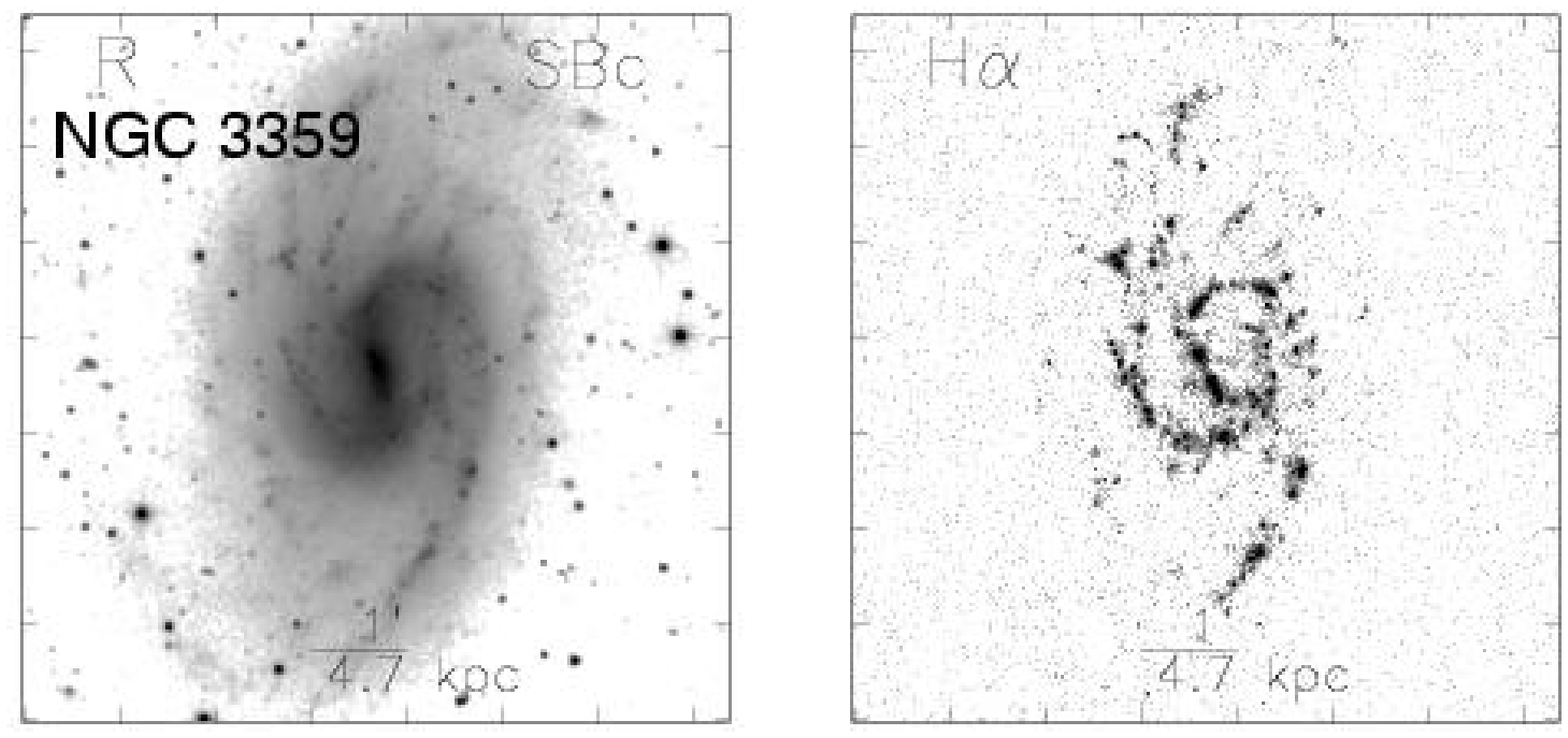}
\includegraphics[scale=0.35]{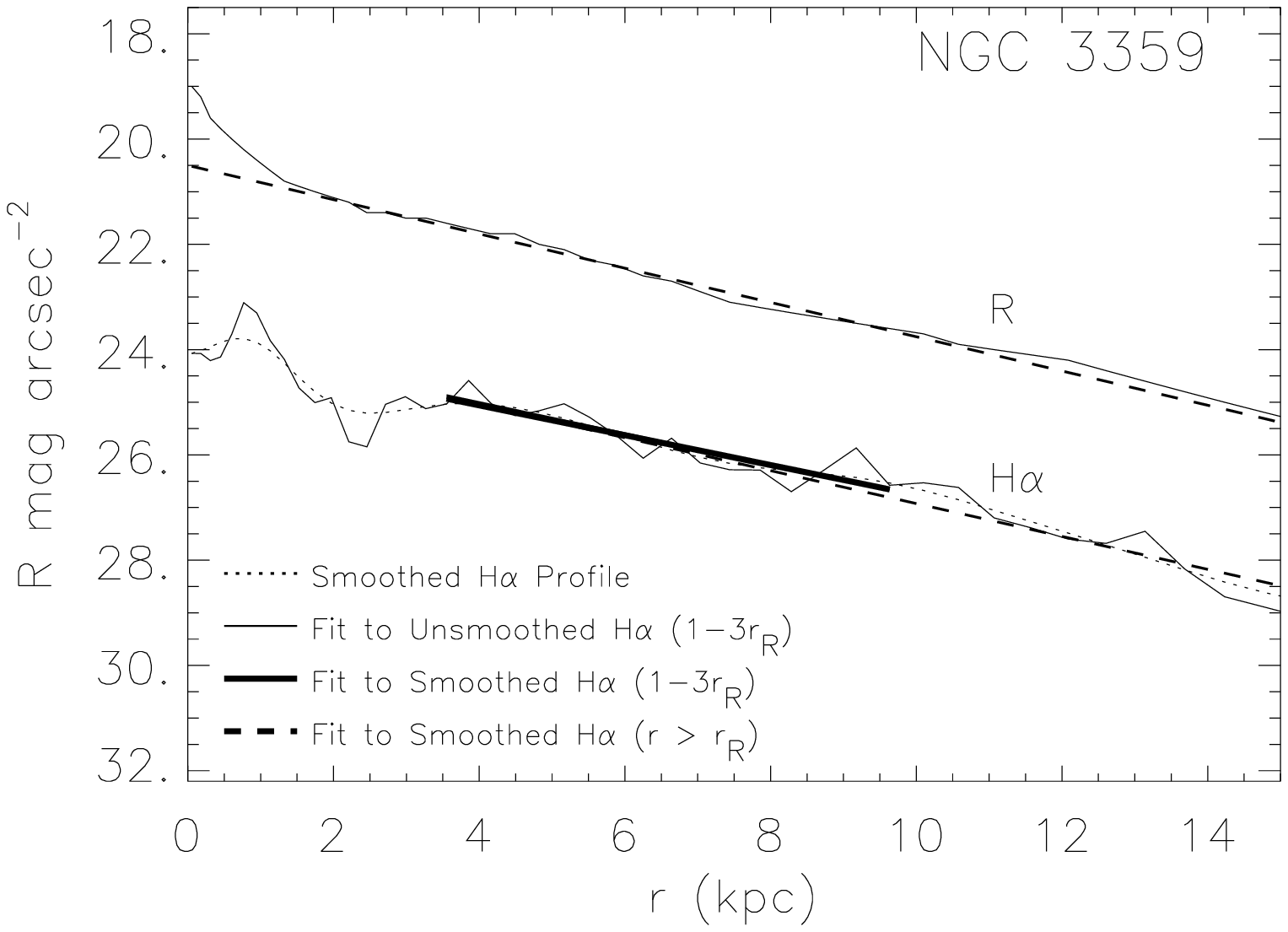}
\includegraphics[scale=0.55]{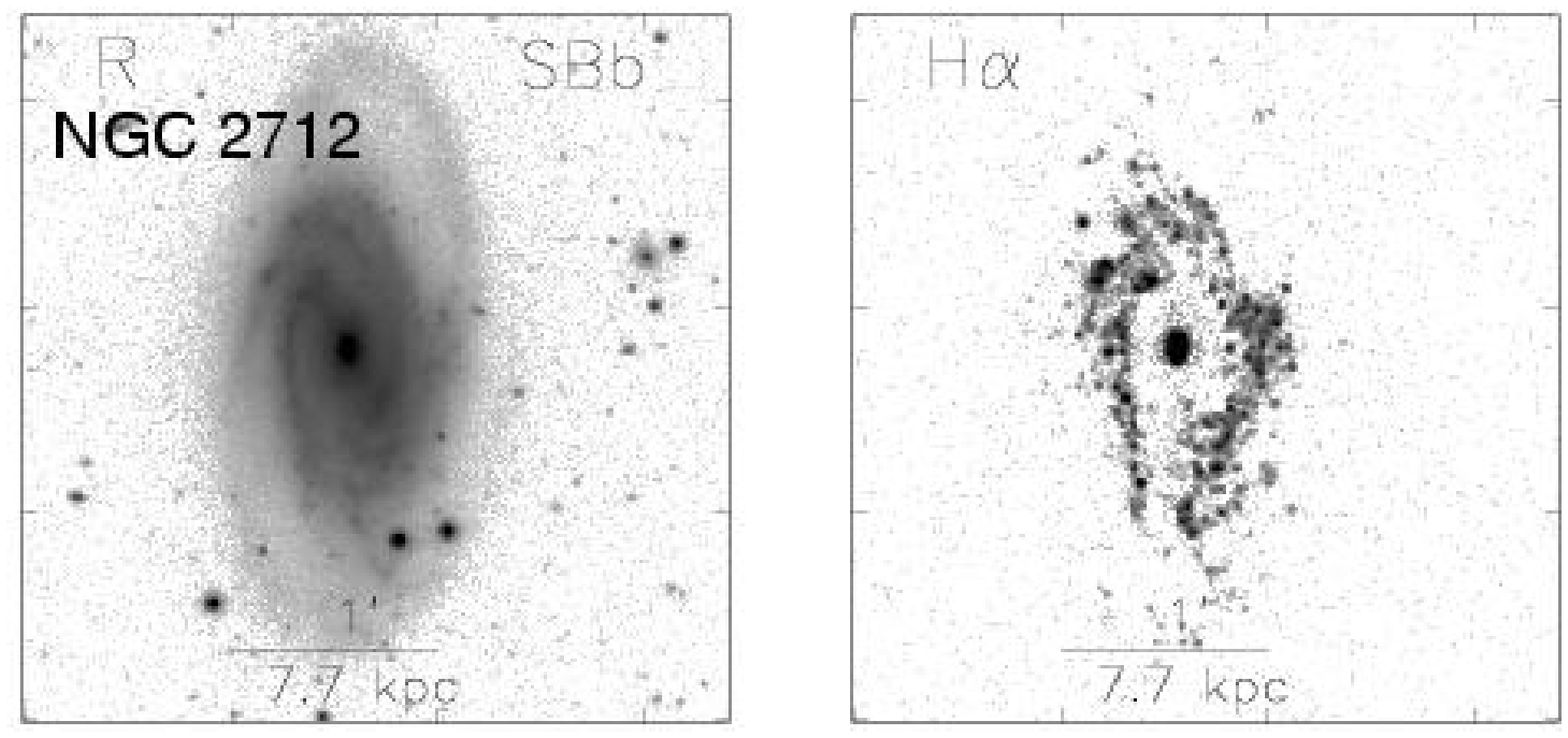}
\includegraphics[scale=0.35]{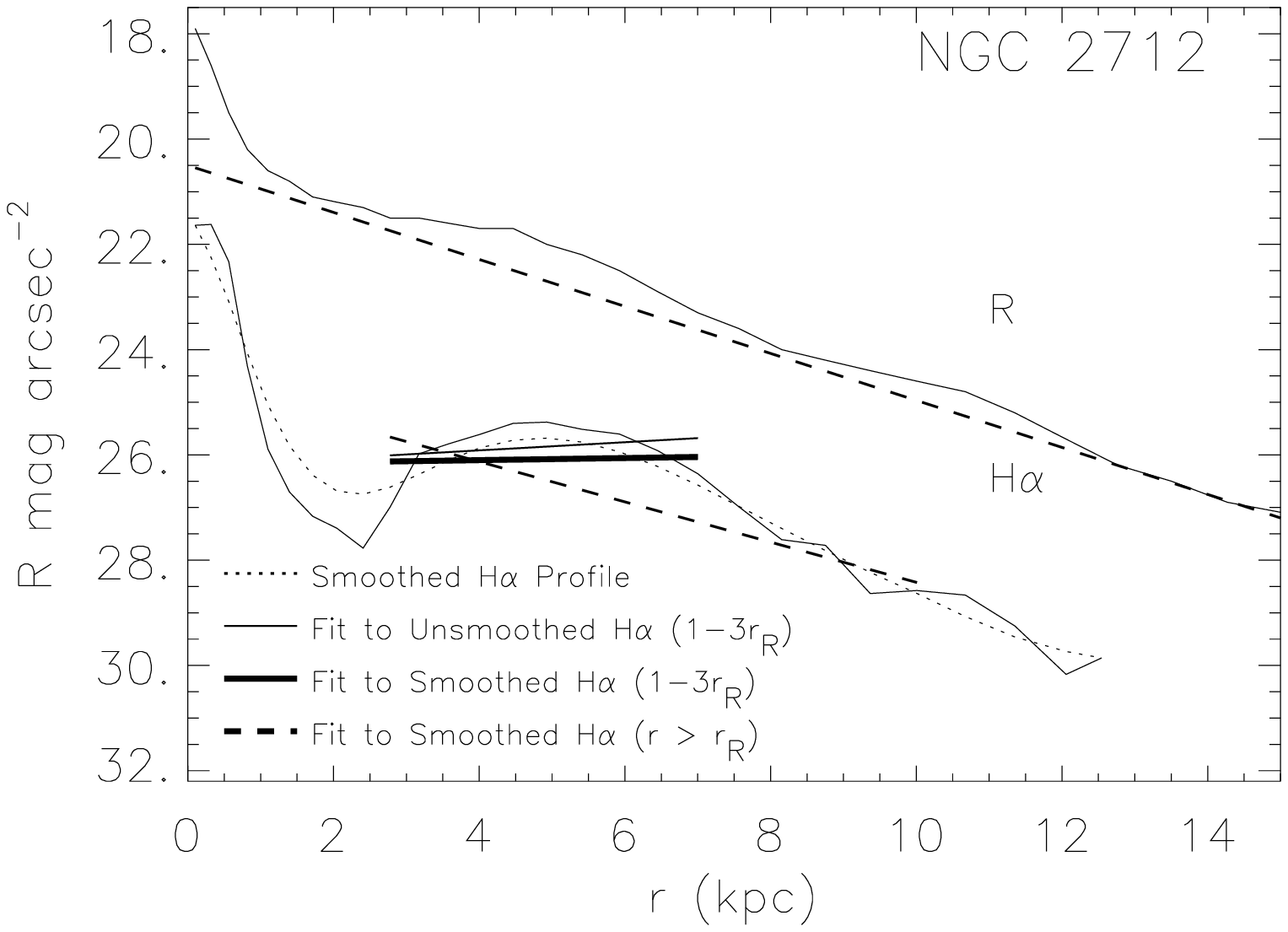}
\caption
{Example smoothed and unsmoothed fits for NGC 3359 (top) and NGC 2712 (bottom).
Scale length fits to the R-band and H$\alpha$ profiles are shown at right.
NGC 3359 has an approximately exponential H$\alpha$ profile and the
H$\alpha$ scale length fits are similar for smoothed and unsmoothed fits over
various radial ranges.
NGC 2712 has a ring of star formation that makes the H$\alpha$ profile
deviate from exponential. Fits to this profile are sensitively dependent on
the degree of smoothing and the radial fitting range adopted.
Galaxies such as this one were eliminated from the analyses in this
paper.
\label{examfits}}
\end{figure*}

The results of the disk scale length measurements described in
Section~\ref{sfit} are presented in Table~\ref{tabprop}, along with a
few quantities that will be used in the subsequent analysis. The
galaxies are ordered by increasing Right Ascension. The parameters listed are:

\noindent
{\bf Col. 1:} NGC or IC designation. \\
{\bf Col. 2:} other name, from the UGC (Nilson 1973), ESO (Lauberts 1982), 
or MCG (Vorontsov-Velyaminov \& Arhipova 1968) catalogs. MCG
designations are abbreviated to eight characters. \\
{\bf Col. 3:} RSA morphological type (from Sandage \& Tammann 1987, or 
Binggeli \etal ~1985, or Sandage \& Bedke 1994). We list
here only the basic type, i.e. we drop from the name luminosity class and 
other details.\\
{\bf Col. 4:} RC3 morphological type (from de Vaucouleurs \etal
~1991). Again, only the basic type is provided. \\
{\bf Col. 5:} B$_T^0$, total, face-on blue magnitude from the RC3 catalog. \\
{\bf Col. 6:} b/a, galaxy axial ratio used for the surface photometry, 
and, in parentheses, inclination to the line of sight in degrees, calculated 
using the Hubble (1926) conversion. \\ 
{\bf Col. 7:} distance, in Mpc. A mean Virgo distance of 16.6 Mpc is 
assigned to all
cluster members; for field objects, the distance is calculated from the Virgo
infall model of Tonry \etal ~(2000). \\
{\bf Col. 8:} \hicont, HI content parameter, calculated following the 
prescription of Giovanelli \& Haynes (1984; see Section~\ref{HI}). \\
{\bf Col. 9:} $r_{H\alpha95}$, radius containing 95\% of the total 
H$\alpha$ flux  (Section~\ref{rmax}), in arcseconds. \\
{\bf Col. 10:} $r_R$, R disk scale length, and corresponding error, 
in arcseconds. \\
{\bf Cols. 11-13:} H$\alpha$ disk scale lengths measured in the
radial intervals 1-3$r_R$, 1-2$r_R$, and $r_R-r_s$, respectively, and
corresponding errors, in arcseconds. Entries are given only for those
objects for which satisfactory fits could be obtained. Scale lengths 
with measurement errors larger than 20\% are marked with a dagger. 
Galaxies with sufficient emission over the given fitting range, but 
for which the fit produced a negative or zero value (clearly
unphysical) are marked with an asterisk (to distinguish them from cases where a
measurement of the scale length could not be obtained due to lack of
H$\alpha$ emission in the radial range considered). 
An example of this situation was 
presented in Figure~\ref{examfits}, where the exponential fit to the H$\alpha$
profile of NGC 2712 over the 1-3$r_R$ interval has a positive slope.\\
{\bf Col. 14:} $r_{max}$, maximum extent of the H$\alpha$ rotation
curve (Section~\ref{rmax}), in arcseconds.

R--band and H$\alpha$ surface brightness profiles are depicted for all
sample galaxies for which an H$\alpha$ scale length could be determined
in the figures in the Appendix. For display purposes
we divided the sample into: 
(a) 34 field galaxies and 19 Virgo cluster members with H$\alpha$ 
emission that can be traced to r$\geq 3r_R$  and have exponential profiles 
over 1-3$r_R$ (Figures \ref{scalefig} and \ref{scalefigv}, respectively); and
(b) 13 field and 14 Virgo cluster galaxies with H$\alpha$ emission that
could not be fit within 1-3$r_R$ because their profiles were either not
sufficiently exponential (see Figure~\ref{examfits}) or were less
extended than $3r_R$ (Figures \ref{scalefigtr} and \ref{scalefigtrv},
respectively). The remaining galaxies (4 in the field sample, 19 in
the Virgo cluster) do not have entries in columns 11-13 of
Table~\ref{tabprop} and are not shown.
The H$\alpha$ profiles are displaced by arbitrary amounts for clarity;
the radial coordinate is expressed both in kpc (top of each panel) and
in units of the B-band isophotal radius $r_{25}$. Profiles are plotted
out to the radius containing the outermost resolved HII region.
Exponential fits are 
overplotted on the R--band and H$\alpha$ profiles; in the latter case
the fitting ranges adopted are also shown as solid (1-3$r_R$, Figures
\ref{scalefig} and \ref{scalefigv}) or dashed ($r_R-r_s$, Figures
\ref{scalefig}--\ref{scalefigtrv}) lines.
As mentioned in Section~\ref{hascale}, $r_{\alpha}(r_R-r_s)$ scale
lengths are obtained from exponential fits to the H$\alpha$ profiles
that include only points with an uncertainty
smaller than the sky noise (whereas this restriction does not apply to
the two other definitions of H$\alpha$ scale length adopted in this
work). As a result, the corresponding dashed lines in the figures in
the Appendix do not always reach the outermost points of the H$\alpha$
profiles, and, in the case of Figures \ref{scalefig} and \ref{scalefigv},
they are sometimes even shorter than the solid lines (traced over the
1-3$r_R$ interval).

A quick inspection of Table \ref{tabprop} reveals a first result: despite
the fact that the luminosity and morphological type distributions of the 
Virgo and field samples
are comparable, a much larger fraction of the Virgo galaxies does
not exhibit star formation over the outer disk. Because of the absence of 
HII regions in their outer disks, we are obviously unable to measure
H$\alpha$ scale lengths for those galaxies with truncated star
forming disks. 

\begin{figure}
\includegraphics[scale=0.5]{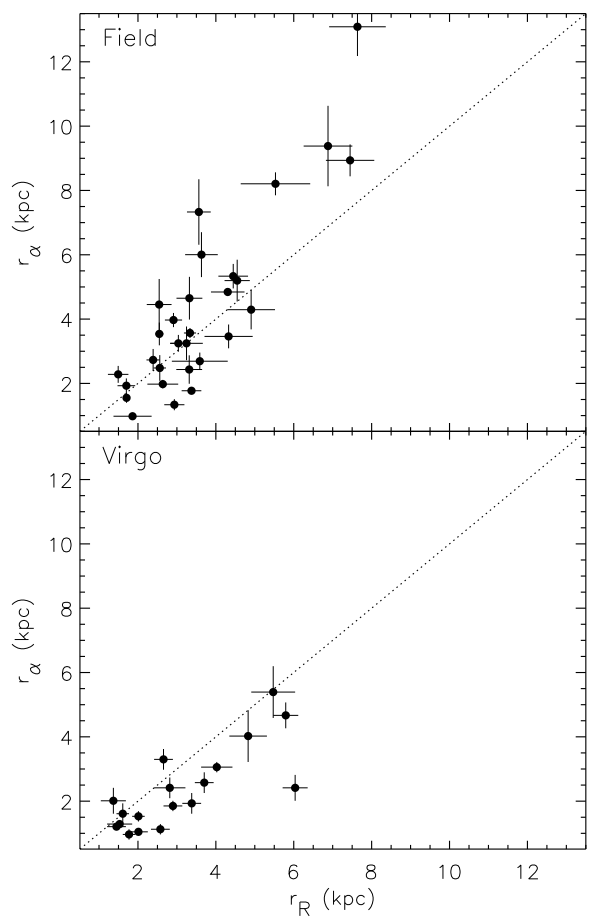}
\caption
{H$\alpha$ scale length as a function of R-band scale length in kpc, 
separately, in the field (upper)
and Virgo Cluster (lower) environments for galaxies with 
fitting errors of 20\% or less. The dotted line shows a one-to-one 
relationship. H$\alpha$ scale lengths of Virgo galaxies are significantly
shorter than the R-band scale lengths.
\label{plotslwsa}}
\end{figure}

\begin{figure}
\includegraphics[scale=0.5]{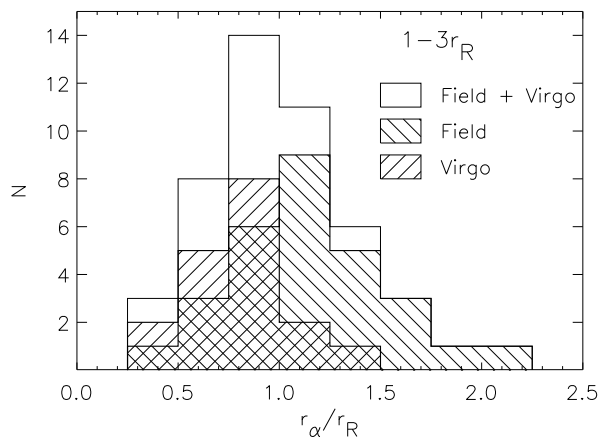}
\caption
{Histograms showing the distribution of the H$\alpha$/R-band scale
length ratio for galaxies in the field and Virgo (oppositely hatched)
and total (clear) samples. H$\alpha$ scale lengths are obtained from
exponential fits to the surface brightness profiles over the 1-3$r_R$
range; only measurements with a fitting error of 20\% or less are included.
The field and Virgo samples are different at a level of 99\%, according to a
K-S test. In the field, the H$\alpha$ scale lengths are on average about
14\% longer than R-band ones, but with large scatter, while for the Virgo
sample, H$\alpha$ scale lengths are 20\% shorter (see Table~\ref{fitresults}).
\label{histwsa}}
\end{figure}

\begin{figure}
\includegraphics[scale=0.5]{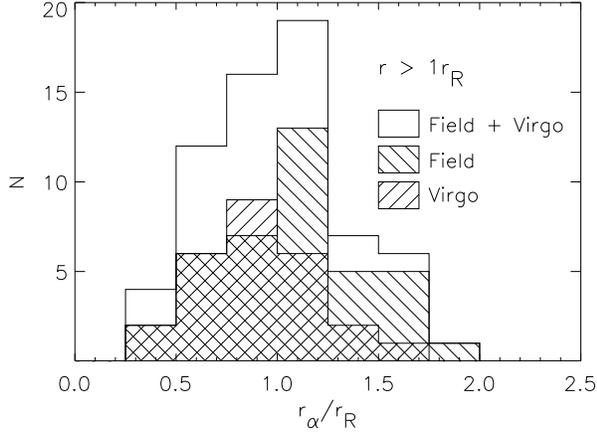}
\caption
{Same as Figure~\ref{histwsa} for H$\alpha$ scale lengths measured
over the radial range $r_R-r_s$. The average scale length ratios for
the field and Virgo samples are similar to those shown in
Figure~\ref{histwsa} (see also Table~\ref{fitresults})
\label{histall}}
\end{figure}

A detailed comparison between the R--band and the various types of
H$\alpha$ scale lengths defined in Section~\ref{hascale} has been
made using measurements with fitting errors of 20\% or less. Our
results are summarized in Table~\ref{fitresults} for the field and
Virgo samples separately. In the case of H$\alpha$ scale lengths
measured over $r_R-r_s$, we show also the results obtained using the
whole samples, without error constraints.
As Table~\ref{fitresults} clearly shows, the $r_{\alpha}/r_R$ scale
length ratio is significantly different for the field and cluster
galaxies: regardless of the method used to measure $r_{\alpha}$, the
H$\alpha$ emission is on average more extended than that traced in the
R--band for the field sample, and less extended for the Virgo members.
This is illustrated in more detail in
Figures~\ref{plotslwsa}-\ref{histall}, as discussed below.

\begin{deluxetable}{lccrr}
\tabletypesize{\scriptsize}
\tablecaption{Scale Length Fits}\label{fits}
\tablecolumns{5} 
\tablewidth{0pt}
\tablehead{
\colhead{Fitting Range}& 
\colhead{\# of}&
\colhead{\% of}&
\colhead{Average}&
\colhead{$\sigma$}\\
\colhead{}& 
\colhead{Galaxies}&
\colhead{Sample}&
\colhead{r$_{\alpha}$/r$_R$}&
\colhead{}\\
\cline{1-5}\\[-3pt]
\multicolumn{5}{c}{\bf Field Results}
}
\startdata
1-3$r_R$ ($<$20\% err)& 29 & 57 &1.14 $\pm$ 0.07 & 0.39\\
1-2$r_R$ ($<$20\% err)& 24 & 47 &1.18 $\pm$ 0.10 & 0.48 \\
$r_R-r_s$ ($<$20\% err)&39 & 76 & 1.06 $\pm$ 0.06  &0.36\\
$r_R-r_s$  & 48 & 94 & 1.20 $\pm$ 0.09 &0.65\\

\cutinhead{\bf Virgo Results}
1-3$r_R$ ($<$20\% err)& 18 & 35 &0.79 $\pm$ 0.06 &0.27\\
1-2$r_R$ ($<$20\% err)& 20 & 38 &0.91 $\pm$ 0.05 &0.23\\
$r_R-r_s$ ($<$20\% err)&26 & 50 &0.86 $\pm$ 0.06& 0.29\\
$r_R-r_s$ & 35 & 67 & 0.88 $\pm$ 0.07 & 0.39\\
\enddata
\label{fitresults}
\end{deluxetable}

Figure~\ref{plotslwsa} shows the relationship between R--band and H$\alpha$ 
scale lengths, obtained from exponential fits to the H$\alpha$
profiles over 1-3$r_R$, for the two samples separately. For the field
sample (upper),
there is some evidence that galaxies with larger stellar scale lengths have
proportionately larger H$\alpha$ ones. For Virgo galaxies (lower), it is clear
that the star formation is more centrally concentrated (smaller $r_{\alpha}$)
than the stellar disk is. As seen in Table~\ref{fitresults}, H$\alpha$
scale lengths are, on average, 14\% longer than the R--band ones in
field galaxies, and 20\% shorter in cluster objects.
Figure~\ref{histwsa} provides a histogram representation of the same data
subset. The scatter for both samples is large, but the difference between the 
field and
Virgo sample is significant at the 99.8\% level according to a K-S test.
When the Virgo and field samples are combined, the H$\alpha$ and
R--band scale lengths are the same on average. The results show no significant
difference if either the unsmoothed or the smoothed H$\alpha$ profiles
are adopted for the fitting process. 

As noted in Section~\ref{hascale} and evident in Figures~\ref{scalefig} and
~\ref{scalefigv}, some galaxies with HII regions beyond 3$r_R$ 
have H$\alpha$ surface brightness \it within 3$r_R$ \rm
comparable or below that of the sky. The $r_R-r_s$ fits can be used to test
the effect of including these galaxies in the 1-3$r_R$ fitting results.
The difference in 1-3$r_R$ and $r_R-r_s$ fitted scale lengths is greater 
than the uncertainties
in the fits for 1 field galaxy (NGC~2805) and 5 Virgo galaxies 
(NGC~4192, NGC~4321, NGC~4501, NGC~4535, and NGC~4639). If these galaxies
are eliminated from the sample, the results are not  
significantly
different (Field: 1.13$\pm$ 0.07; Virgo: 0.81$\pm$ 0.08). 

It is also interesting to further consider H$\alpha$ scale lengths determined
from exponential fits to all points in the surface brightness profiles
beyond $r_R$ that had an uncertainty less than the sky uncertainty. 
This fitting method is in fact more similar to that of RD94, 
who made their fits to the outer disk and did not adopt a radial range. 
Figure~\ref{histall} is analogous to Figure~\ref{histwsa}, except for
the H$\alpha$ scale length adopted. The comparison between these two
figures shows that the results for the larger fitting range are similar
to those of the more limited radial range (see also 
Table~\ref{fitresults}). The difference between the field and Virgo
samples is at the 92\% level.
If we include all 83 galaxies without an error cut (maximum error 50\%), 
we find similar results with larger dispersion, as given in 
Table~\ref{fitresults}.

Finally, we use the H$\alpha$ scale lengths that correspond to fits in
the 1-2$r_R$ range to investigate the effect of truncation of the outer disk.
We first compare the 1-2$r_R$ and 1-3$r_R$ H$\alpha$ scale lengths.
For an exponential dust-free disk, we would expect these two 
scale lengths to be the same. We tested this for a total of 34
galaxies (21 field and 13 Virgo galaxies) 
with good $r_{\alpha}$ measurements in both intervals, and found that 
the 1-2$r_R$ are, on average, 13\% and 29\% longer than the 1-3$r_R$
ones for the field and Virgo samples, respectively. 
There is no significant difference between the two samples according
to a K-S test.
An inspection of the individual plots presented in Figures
\ref{scalefig} and \ref{scalefigv} reveals that 
several of the Virgo galaxies with the most deviant ratios
have H$\alpha$ profiles that begin to decline sharply between 2 and 
3$r_R$, as a result of the truncation of the disk, therefore leading
to the larger factor for the Virgo sample. 
As we will show in Section~\ref{compare},
there is a strong correlation between the maximum extent of the
star forming region and the H$\alpha$ scale length, which causes a
bias in the scale lengths measured for truncated disks (they are
systematically smaller than those of similar, non-truncated disks).
We also compared $r_{\alpha}$(1-2$r_R$) and R--band scale lengths,
as summarized in Table~\ref{fitresults}.
Note that 1-2$r_R$ scale lengths could not be fit to 
1/3 of the Virgo Cluster sample and two isolated/field galaxies
(NGC 4395 and NGC 4984) that have H$\alpha$ profiles truncated 
within 2$r_R$.
We find again that the Virgo H$\alpha$ scale lengths are 
systematically smaller than the field sample, 
a result that is significant at the 99\% level. 

As a check on the importance of the radial range used to fit
R scale lengths, we refit R scale lengths over the same radial ranges used
to determine the H$\alpha$ ones and recalculated all scale length
ratios. All results were the same within the uncertainty.

In the rest of this paper we will adopt $r_{\alpha}$(1-3$r_R$) as our
measure of H$\alpha$ disk scale length (unless otherwise noted), and
will no longer specify the fitting range.

\subsection{Comparison with Previous Results}\label{previous}

Our results for the field galaxies suggest that H$\alpha$ scale lengths 
are longer than stellar scale lengths, but our value differs from
RD94 significantly. In fact our value of 
$r_{\alpha}/r_R$ = 1.14 $\pm$ 0.07 is closer to the value of unity,
as measured by Hodge \& Kennicutt (1983), than it is to the
$r_{\alpha}/r_V$ = 1.6 $\pm$ 0.1 measured by RD94. 
A Student t-test shows 
differences at the 95\% and $>$ 99\% confidence levels respectively. 
This is surprising, given that our results and those of RD94 are both derived
from CCD images and surface photometry, while those of Hodge \& Kennicutt 
(and Athanassoula \etal ~1993) are derived from the 
different method of HII region 
counting. In this section, we discuss possible reasons for the discrepancy 
between our result and that of RD94.

There are only four galaxies in common between our sample and the RD94 
sample. We agree on the R--band scale lengths for NGC 1637, but are our values
for NGC 4192, NGC 4548, and NGC 6118 are 6-16\% longer 

RD94 are able to measure H$\alpha$ scale lengths for all of these galaxies
except NGC 1637. 
We find the largest H$\alpha$ scale length discrepancy between our
results for NGC 6118. 
While RD94 measure an  H$\alpha$ scale length of 10.7 kpc, we find 
4.2 kpc (1-3$r_R$) or 4.5 ($r_R-r_s$) kpc. We do not reproduce the outer
extent of the H$\alpha$ profile given in Ryder \& Dopita (1993) 
and examination of their image and profile suggests that their outer profile 
may be contaminated by sky. We note that NGC 6118 has the third longest 
H$\alpha$ scale length measured in the RD94 sample. In the
case of NGC 4192, we find a shorter H$\alpha$ scale length of 4.7 kpc 
(1-3$r_R$) or 5.8 kpc ($r_R-r_s$), compared with RD94's value of 
7.8 kpc. Our H$\alpha$ scale lengths are in closer agreement
for NGC 4548, where we find 4.8 kpc ($r_R-r_s$), compared to RD94 5.1 kpc. 

It is hard to draw a general conclusion from the comparison of 3-4 galaxies.
It is perhaps suggestive that we tend to find longer R--band and 
shorter H$\alpha$ scale lengths. 
RD94 note in their paper that their R--band scale
lengths could be systematically low, due to the use of the mode rather than
the median/mean in the derivation of the azimuthally-averaged radial profile.
They find that trial mean surface photometry profiles produced somewhat
longer V-band and I-band scale lengths that were more similar to the 
H$\alpha$ scale 
lengths. As noted above, we do indeed find that our R--band scale lengths are 
longer for 3 galaxies. It is interesting that Athanassoula \etal 
(1993) account for their \it shorter \rm scale length ratios compared to 
Hodge \& Kennicutt (1983) as due to 
systematically longer \it stellar \rm 
scale lengths, rather than shorter
HII region scale lengths. It is therefore possible that the well-known
difficulty of measuring R--band scale lengths contributes to the
discrepancy between our results. 

RD94 did not specify a fitting range, rather they fit the outer disk points
above the noise. 
This does not seem to explain the difference in our results, since
we obtain similar results whether we use the 1-3$r_R$ fitting range or
the $r_R-r_s$ range. We do however note that RD94's spiral sample is
comprised of somewhat later type (by 0.25 in the RC3 type parameter)
and brighter spirals (by 0.4 blue
magnitudes), on average, than even 
our field sample, both of which might lead to some expectation of 
larger H$\alpha$ distributions as suggested in Figure~\ref{galprop}
(see below).

We have 7 galaxies in common with Hodge and Kennicutt (1983). Our H$\alpha$
scale lengths are 
similar, on average, to theirs, but differ individually by factors up to 2.

\section{Comparison with Galaxy Properties and other Disk Constituents}\label{compare}

\begin{figure*}
\centerline{
 \mbox{\includegraphics[scale=0.25]{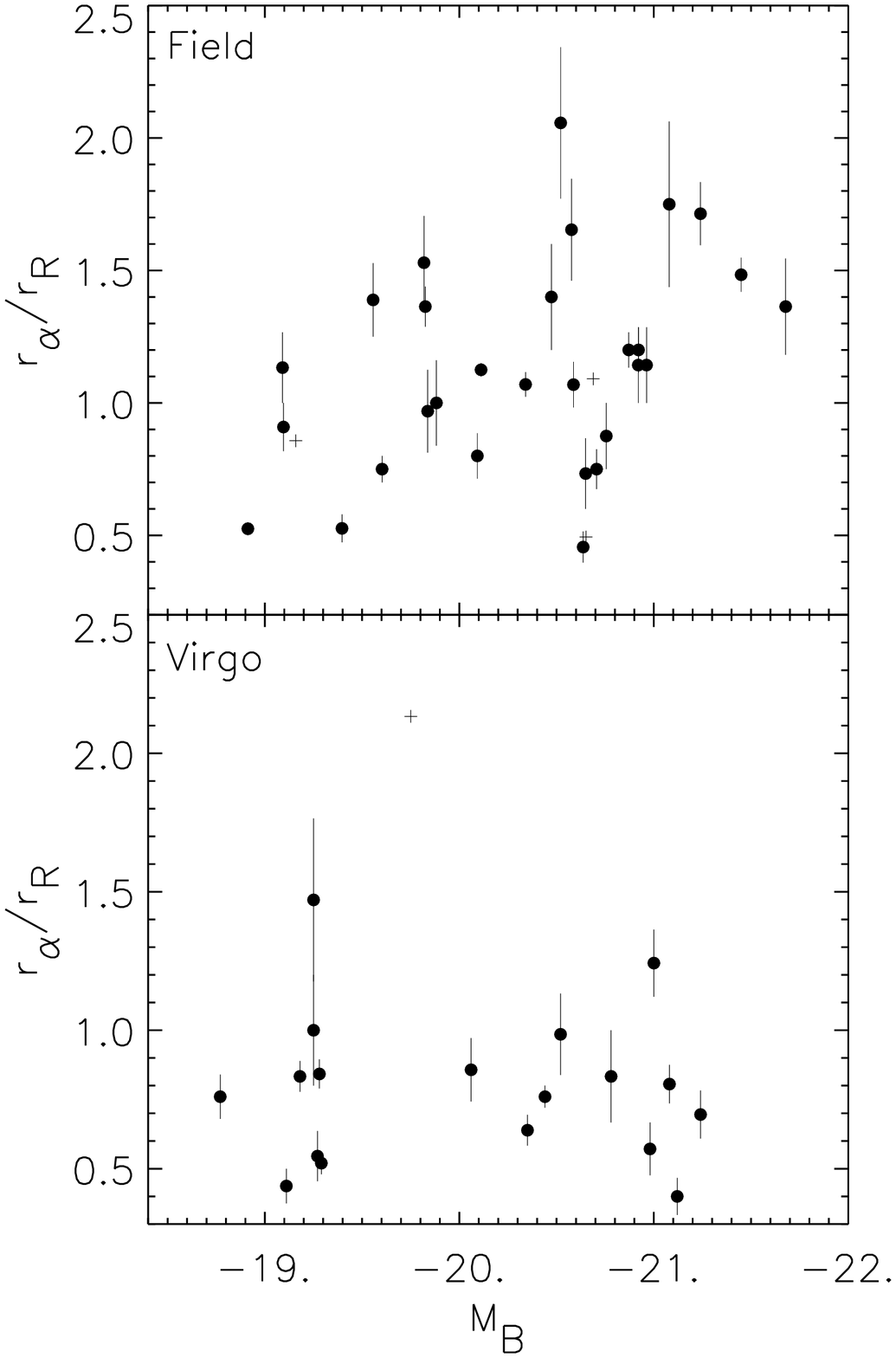}}
 \mbox{\includegraphics[scale=0.25]{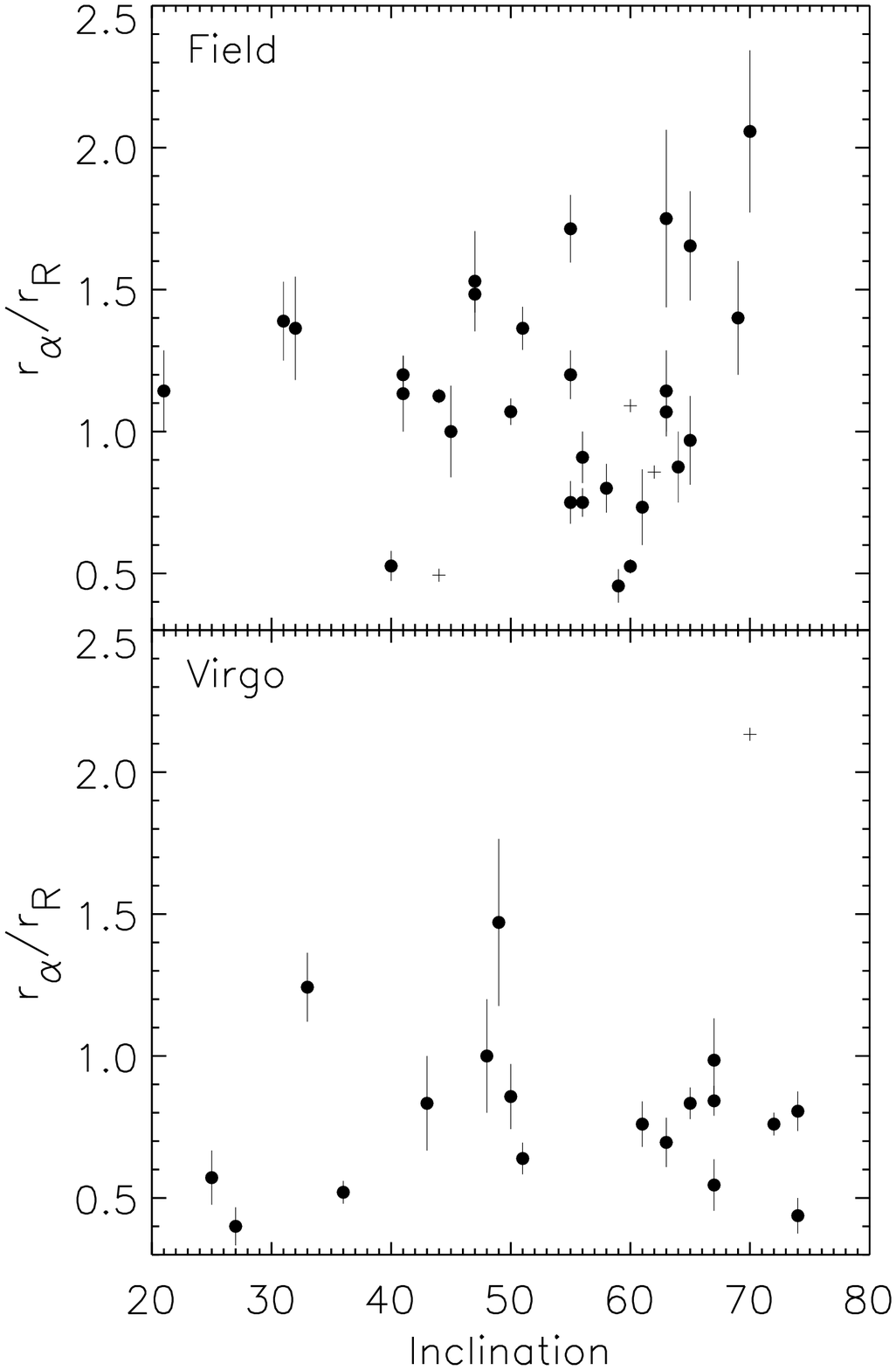}}
 \mbox{\includegraphics[scale=0.25]{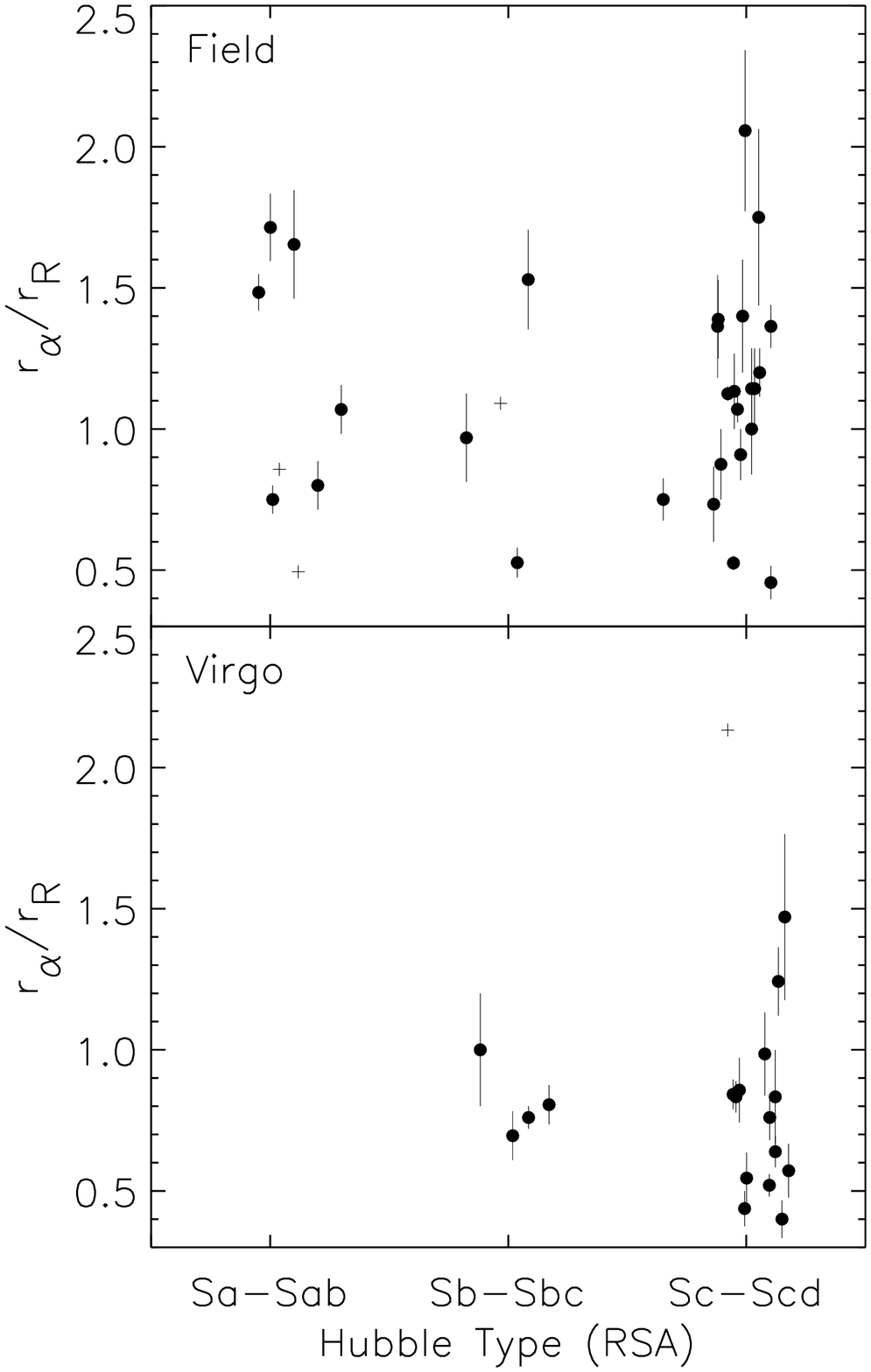}}
 \mbox{\includegraphics[scale=0.25]{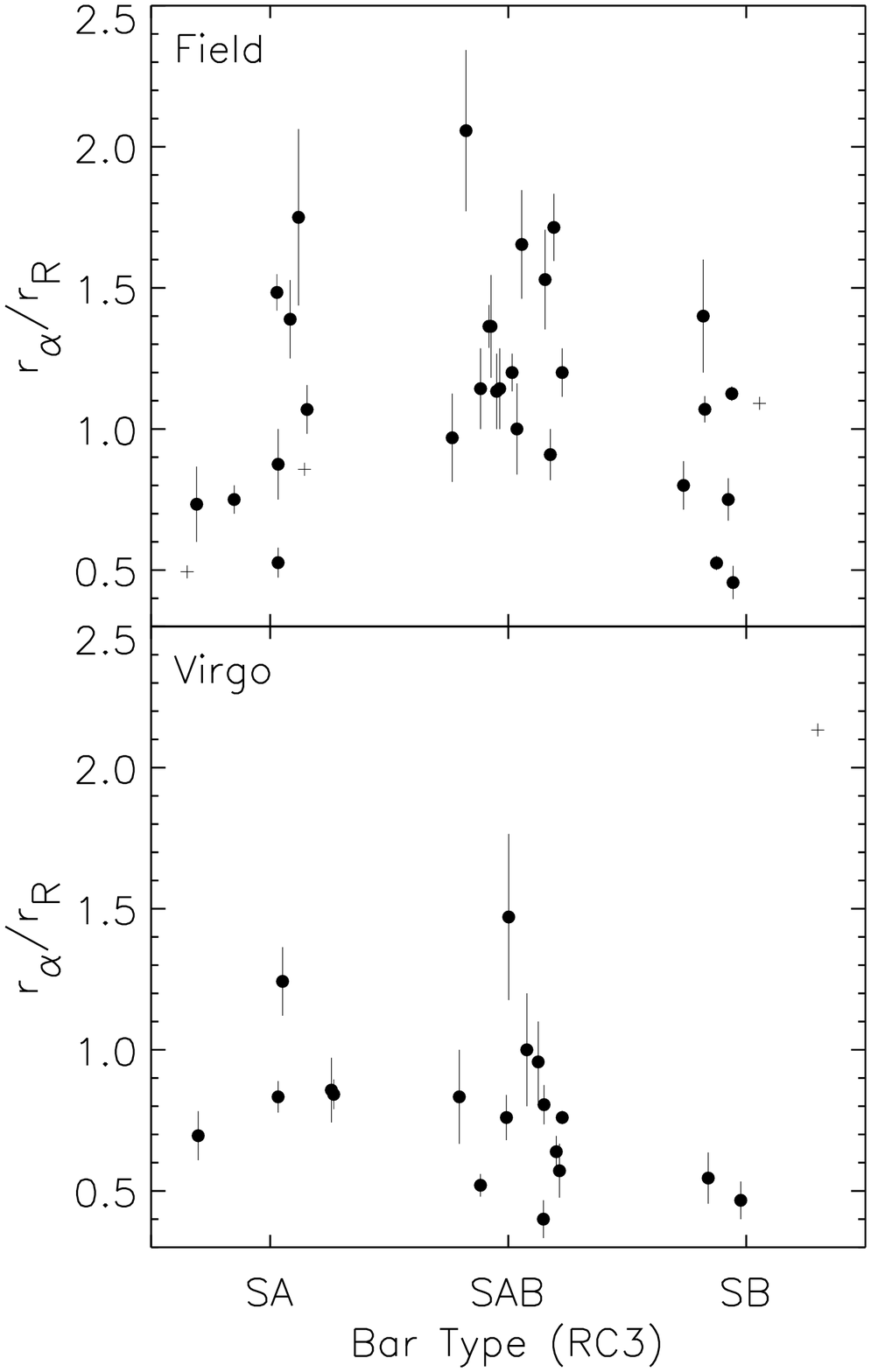}}
}
\caption
{H$\alpha$/R-band scale length ratio as a function of absolute B magnitude,
inclination, Hubble type (RSA), and de Vaucouleurs bar type,
shown separately for
the field (upper) and Virgo cluster (lower) samples. 
The solid symbols
indicate galaxies with uncertainties in H$\alpha$ scale length (measured
over the the 1-3$r_R$ range) $\le$ 20\%. Cross symbols (without
error bars) indicate galaxies with larger uncertainties. \label{galprop}}
\end{figure*}

In this section, we examine how the relative distributions of the older stellar
population as characterized by the R--band scale lengths and the young massive
star forming disk traced by the H$\alpha$ line emission are related to other
galaxy properties and disk constituents. 
We find no strong dependence of H$\alpha$ to R--band scale length ratio
on luminosity, inclination, Hubble type, or de Vaucouleurs bar type, as
shown in Figure~\ref{galprop} for separately the field (upper) and Virgo
cluster (lower) galaxies. We also find no dependence on R--band concentration
parameter, for the more limited sample of galaxies with absolute photometry
(Koopmann \etal ~2001 and Koopmann \& Kenney 2005, provide concentration
values.)
The lack of a dependence on inclination argues that differences in
extinction are not significant, at least in this relatively limited inclination
range (mostly between 40$^\circ$ and 70$^\circ$). Alternatively
the H$\alpha$ and R-band scale lengths could be affected in a similar way by 
extinction.

Although the number of objects included in Figure~\ref{galprop} is quite
small, the figure nonetheless suggests several interesting aspects of the 
present
sample. The leftmost panels suggest a weak tendency for 
brighter field galaxies to have more
extended star forming disks relative to the old stellar population; the same
effect is not evident in the cluster objects, which, except perhaps for the
faintest objects studied here, show systematically small scale 
length ratios. The contrast between the 
number of resultant cluster and field objects of different Hubble
types, when the H$\alpha$ surface 
brightness selection is applied is clear. 
No spirals classified as early types among the Virgo cluster sample 
are characterized
by the necessary star formation rate to yield 
sufficiently extended H$\alpha$ emission. Therefore,
they appear in Table \ref{tabprop}, but are absent from this figure. 
In contrast, in the field, the spirals classified as Sa-Sab do exhibit 
spatially extended star formation.
This result is consistent with models in which the star formation rate in
spirals classified as early-type 
is more strongly quenched by the cluster environment than 
in their later type spiral counterparts, possibly
because they lie on more radial orbits (Dressler 1986; Solanes \etal ~2001).
Note that the Hubble classification itself is likely influenced by the star
formation rate, in that HI deficient spirals with low star formation
rates are more likely to be classified as early-type regardless of their
bulge-to-disk ratio (Koopmann \& Kenney 1998).

\subsection{Comparison with I-Band Scale Lengths}

As a comparison with an even older stellar population
and a confirmation of our surface brightness profile extraction process,
we compare the R--band scale lengths derived above
to ones derived from photometric I--band images
available to us for 23 field and 22 Virgo galaxies
from the  SFI++ Cornell archive (Masters, K. L. \etal, in prep.). On average
$r_R /r_I$ = 1.17 $\pm$ 0.03 (dispersion: 0.20), with no
difference between the field and Virgo samples.
This value is similar to the value of 1.10 $\pm$ 0.02 found by Cunow (2001),
who also found that the ratio increased from 1.0 for face-on to 1.2 for edge-on
galaxies. Other studies (e.g., Peletier \etal ~1994; Cunow 2004) also
find an inclination dependence when comparing galaxies of more extreme 
edge-on and face-on orientations. We find no strong correlations with
inclination, but again, since our sample
is not constructed to sample the extremes of inclination, it is not
sensitive for this test.

Combining our average R:I-band and H$\alpha$:R-band ratios, we find that
the H$\alpha$ scale length is a factor of 1.33 $\pm$ 0.08 (dispersion 0.52) 
longer than I-band scale lengths in field samples. This result compares to 
$r_B /r_I$ of 1.12 for face-on spirals (de Jong 1996b) and
1.36 $\pm$ 0.03 for edge-on spirals (de Grijs 1998). The $r_{\alpha}/r_I$ 
ratios measured in this study are thus very similar to the $r_B /r_I$ ones,
reflective of the similarity of stellar population
sampled by both B-band and H$\alpha$ emission.

\subsection{Comparison with the Maximum H$\alpha$ Extent}
\label{rmax}

As noted above, we find in this study that
Virgo Cluster $r_{\alpha}/r_R$ scale length ratios are systematically 
smaller than those
of field galaxies. To determine if this is due to the smaller size of
Virgo star-forming disks (e.g., Koopmann \& Kenney 2004a),
we also derive estimates of the maximum extent of star formation 
from available imaging and spectroscopic data. 

\begin{figure*}
\includegraphics{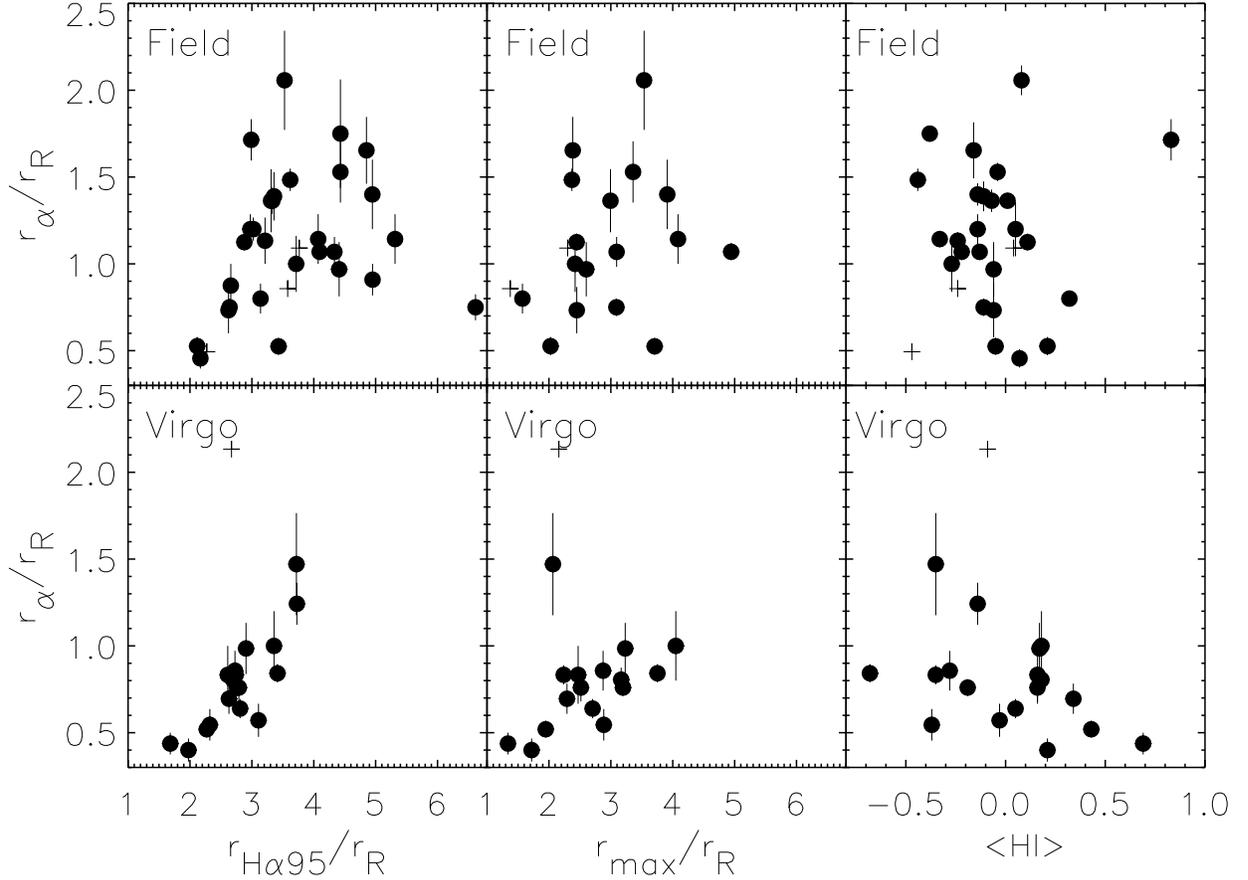}
\caption
{The H$\alpha$/R-band scale length ratio is plotted here as a function of
extent of star formation and HI content, separately for the field (upper)
and Virgo cluster (lower) samples. The solid symbols
indicate galaxies with uncertainties in H$\alpha$ scale length (measured
over the the 1-3$r_R$ range) $\le$ 20\%. Cross symbols (without
error bars) indicate galaxies with larger uncertainties. 
The star formation extent is
characterized by $r_{H\alpha95}$, the radius encompassing 95\%
of the H$\alpha$ emission (left panels), or $r_{max}$, the outermost
point of the folded H$\alpha$ rotation curve (middle), normalized by
the R-band scale length. $r_{\alpha}/r_R$ is strongly correlated with the
size of the star forming disk in the Virgo sample.
The panels on the right show $r_{\alpha}/r_R$ as a function of HI
content (see text). Galaxies with $\langle {\rm HI} \rangle \geq 0.5$
are considered to be HI--poor. There is a weak trend for Virgo galaxies
with less HI content to have less extended star-forming disks.
\label{plot95}}
\end{figure*}

First, using the imaging data, we characterize the size of the H$\alpha$ disk 
as $r_{H\alpha95}$, the radius containing 95\% of the H$\alpha$ emission 
(see Koopmann \etal ~2001).  The results are shown in the leftmost panels
of Figure~\ref{plot95}. There is a large scatter, but 
the H$\alpha$ scale lengths in the Virgo Cluster sample
are systematically shorter for galaxies with smaller H$\alpha$ disks
(smaller $r_{H\alpha95}$). The correlation is weaker in the field sample. 
Again, we emphasize that, unlike the field
population, there are {\it no} Virgo galaxies with ratios of
$r_{H\alpha95}/r_R > 4$, whereas a significant fraction of the
field galaxies show star formation extending well out into the stellar
disk. 

As a second comparative test, we also exploit a measure of
the maximum extent of the H$\alpha$ rotation 
curve, $r_{max}$, making use of the high sensitivity long-slit optical
rotation curves described in Catinella \etal ~(2005). Following those
authors, $r_{max}$ is simply the radius of the outermost radial
point with detectable H$\alpha$ emission, as measured from the folded 
rotation curve. As evident in the central panels of
Figure~\ref{plot95}, this measure also reveals the truncation of
star formation in Virgo spirals, confirming the result found from
the imaging dataset.

These results are somewhat surprising because the Virgo galaxies 
with H$\alpha$ scale lengths measured in the 1-3$r_R$ range
are among the most `normal'
in the cluster - they have normal HI contents (Section ~\ref{HI}) 
and their disks are not
severely truncated. While even `normal' Virgo galaxies do show a
tendency to smaller H$\alpha$ disks outside of r$_{24}$ in R (as can
be seen in Figure 2 of Koopmann \& Kenney 2004b), this analysis considers
only the star formation rates within 3$r_R \sim r_{24}$. 
This suggests that the reduction in H$\alpha$ scale length is due not only to a
simple truncation in the star-forming disk, but also to an increase in
the central concentration of star formation.
Evidence that this is the case can be seen
in Figures \ref{plotslwsa} and \ref{plot95}: the stellar disks of both Virgo
and field galaxies span a similar $r_R$ range, but the Virgo HII disks
are clearly truncated, exhibiting not just a cutoff in the extent of
the star forming regions, but also a change in the slope of the
exponential decline of the H$\alpha$ surface brightness. 
In fact, Koopmann \& Kenney (2004a) show that Virgo spiral galaxies
tend to have normal -- enhanced inner disk (0-0.3r$_{24}$) star formation 
rates compared to isolated counterparts. We calculated the star formation 
rates over the 0-0.3r$_{24}$ range for our sample of Virgo Sc-Scd and found 
that these rates are 30\% higher in the median than those of isolated Sc-Scd 
galaxies (as given in Koopmann \& Kenney 2004a). 
Thus star formation in Virgo spirals 
is quenched by processes that affect the disk from the outside in, 
and act in at least some cases to increase inner disk
star formation rates, yielding more centrally concentrated star
formation compared to similar galaxies found in the field. 

\subsection{Comparison with HI content}\label{HI}
Several studies have shown a dependence of star formation rates and
extent on the HI content (e.g., Gavazzi \etal ~2002; Koopmann \&
Kenney 2004a; Catinella \etal ~2005), and it is well known that Virgo
cluster spirals are relatively gas-poor compared to their field
counterparts.  Here, we reexamine this relationship comparing the
field and cluster samples, but first note that the majority of the
Virgo galaxies for which the H$\alpha$ emission is both strong and
extended enough for a scale length determination are relatively
HI-normal. Hence, rather than concentrating on the degree of HI
deficiency, (e.g., Solanes \etal ~2002), we concentrate on the
relative HI content of the gas-bearing population. The HI content
parameter \hicont\ tabulated in Table \ref{tabprop}, adopts the
commonly-used definition of relative HI content which compares the
observed HI mass for a given galaxy with that expected for the
average, isolated galaxy of similar optical linear size and
morphological type.  Note that this definition of \hicont\ corresponds
to the ``HI deficiency'' parameter (Haynes \& Giovanelli 1984);
positive values indicate relative gas poorness.  Here, we use HI
fluxes from the digital compilation of Springob \etal ~(2005) where
available, or else from the literature.  Values of \hicont\ were
calculated following the relations given in Solanes \etal ~(1996) (and
as described in Koopmann \& Kenney 2004a).  Indeed, there is only one
object in each sample depicted in Figure~\ref{plot95} that is truly HI
deficient, with \hicont\ greater than 0.5, that is, containing less
than 1/3 of the HI mass expected for a ``normal'', isolated galaxy of
comparable linear diameter and morphological type.  As evident in
Table \ref{tabprop}, many more Virgo spirals are strongly HI
deficient, but their H$\alpha$ extents are too truncated to permit
even the determination of $r_{\alpha}$.  At the same time, the right
panels of Figure~\ref{plot95} exhibit a weak trend of smaller
$r_{\alpha}/r_R$ ratios with decreasing HI content (larger
\hicont\ values. Notice that since both \hicont\ and $r_{\alpha}/r_R$ are
normalized quantities, this correlation is not a simple scaling
relation).  In other words, at fixed R-band disk scale length,
galaxies with smaller HI content have, on average, less extended
on-going star formation, a result that holds for both field and Virgo
objects that are {\em not} HI-deficient.  This is also in agreement
with the results of Catinella \etal ~(2005), who find, for a much
larger sample of some 700 HI-normal objects, that galaxies with higher
than average HI content have larger than average H$\alpha$ extent, and
vice versa. In conclusion, these results are consistent with previous
findings of H$\alpha$ truncation in HI-stripped cluster galaxies
(Gavazzi \etal ~2002; Koopmann \& Kenney 2004a), and show that the
correlation between massive star formation as traced by H$\alpha$
emission and HI content holds also for HI-normal spirals and is not
limited to the cluster environment.

Vogt \etal ~(2004) find that HI-deficient galaxies in clusters
are offset in the fundamental plane, consistent with a 25\% decrease
in I--band scale lengths. We directly
compared our R--band scale lengths for HI
normal and HI deficient spirals in the Virgo Cluster, but find no
significant difference.

\subsection{Comparison with the Scale Length of Molecular Gas}
\begin{figure}
\includegraphics[scale=0.5]{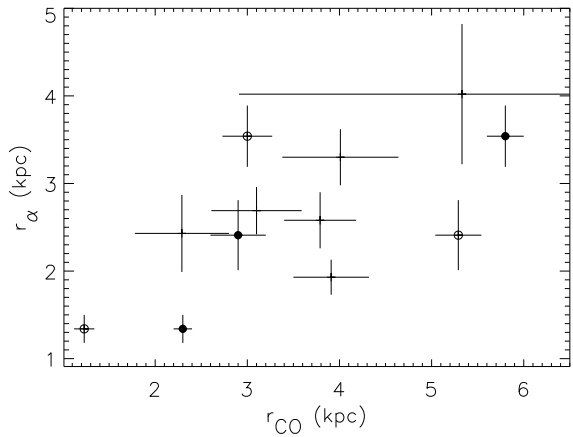}
\caption
{Comparison between H$\alpha$ and CO scale lengths.
CO scale lengths are extracted from Regan \etal~(2001) and Nishiyama 
\etal~(2001). Individual values of $r_{CO}$ measured by independent 
groups can deviate far 
more than the given error bar. This is illustrated using three galaxies in
common to both studies, indicated with filled (Regan \etal) and open 
(Nishiyama \etal) circles. These galaxies are, in order of increasing 
$r_{\alpha}$: NGC 2903, NGC 4321, and NGC 628. 
The H$\alpha$ scale lengths are shorter than the CO ones. This 
surprising result is most likely biased by the inclusion of Virgo 
Cluster galaxies in the Regan \etal ~and Nishiyama \etal ~samples.
\label{haco}}
\end{figure}

Scale lengths measured from CO radial profiles, $r_{CO}$,
were compiled from the studies of  Regan \etal ~(2001) and 
Nishiyama \etal ~(2001) for 9 sample galaxies (4 field and 5 Virgo)
with measured 1-3$r_R$ H$\alpha$ scale lengths.
Two of the field galaxies and one of the Virgo
galaxies are measured in both studies. The scatter
between sources is large, e.g., NGC 4321 has values of 3.89 kpc (Regan
\etal ~2001) and 5.29 kpc (Nishiyama \etal ~2001).
There is a correlation between the H$\alpha$ and CO scale lengths, but
large scatter, as shown in Figure~\ref{haco}.
In the mean 
$r_{\alpha}$/$r_{CO}$ = 0.79 $\pm$ 0.08 (dispersion 0.24). 
This result appears to be somewhat 
biased by shorter Virgo H$\alpha$ scale lengths,
since the average for the Virgo galaxies is 0.67 $\pm$ 0.07, while that
of field galaxies is 0.90 $\pm$ 0.13. 
Regan \etal ~(2001) find that CO and 
stellar disk (K) scale lengths for a sample of 15, mostly field galaxies are 
correlated, with a mean ratio of 0.88 $\pm$ 0.14 (dispersion 0.52). 
Combining this result and our own result for the field, we would expect 
the H$\alpha$ to
CO ratio to be closer to 1.5 - 1.6, taking into account the conversion
between I and K scale lengths (e.g, de Jong 1996b).

\subsection{Comparison with the Scale Length of Dust}

Thomas \etal ~(2004) find that the ratio of the dust scale length, as measured
at 850$\mu$m, to the
blue isophotal diameter has a mean value of 0.33 $\pm$ 0.04 for a sample of
29 near-infrared bright spiral galaxies imaged in the SCUBA Local Universe 
Galaxy Survey  (Dunne \etal ~2000). 
We find a similar value of 0.28 $\pm$ 0.02 (dispersion 0.11) for the ratio of
the H$\alpha$ scale length to the blue isophotal diameter for the field sample.
In the Virgo sample, the ratio is 0.20 $\pm$ 0.03 (dispersion 0.12). 
Xilouris \etal ~(1999) derive dust scale lengths for 7 galaxies,
using broadband visible and near-infrared (I--band) imaging and a model
which assumes that the dust is smoothly distributed within a 3-D axisymmetric 
exponential disk. The resulting dust scale 
lengths are a factor 1.4 $\pm$ 0.2 longer than the stellar scale lengths.
This is similar to our result of $r_{\alpha}$/$r_I$= 1.33 $\pm$ 0.08.
Thus we conclude that H$\alpha$ seems to be distributed similarly to 
the dust in field spirals.

\section{Discussion}\label{summ}

We have compared exponential scale lengths of the stellar and massive
star-forming populations for a large sample of field and Virgo Cluster
galaxies. We find that exponential scale lengths of massive star
formation, as traced by H$\alpha$, are correlated with stellar
exponential scale lengths. In the field, H$\alpha$ scale lengths are
on average a factor of 1.14 $\pm$ 0.07 longer than those measured in
R--band and their ratio is not strongly dependent on
luminosity, inclination, morphological type, central R--band light
concentration, or bar type.  For these field spirals, H$\alpha$
appears to be distributed in a similar way as broadband B and dust
emission, as measured at 850 $\mu$m and near-infrared wavelengths.

Scale length ratios are strongly dependent on the environment of the
galaxy.  The H$\alpha$ scale lengths of Virgo Cluster spiral galaxies
are typically 20\% smaller than R--band scale lengths, and their
ratio is strongly correlated with the size of the star-forming disk.  
In at least some cases, an increased central concentration of star
formation has contributed to the decreased H$\alpha$ scale length.

Environmental processes that affect the gas content of galaxies are
likely responsible for the shortening of the H$\alpha$ scale length in
Virgo cluster galaxies.  Modeling of these processes has become more
realistic in recent years and may provide some insight into our
observed results.

Intracluster medium - interstellar medium (ICM-ISM) interactions (Gunn
\& Gott 1972; Nulsen 1982; Vollmer \etal ~2001; van Gorkom 2004)
remove gas from galaxies that pass through the intracluster medium of
the cluster.  Ram pressure stripping causes truncation of the gas disk
(e.g., Abadi \etal ~1999) and, in the case of high ICM density, can
remove much of a galaxy's gas within 10-100 Myr (Abadi \etal ~1999;
Quilis \etal ~2000; Roediger \& Hensler 2004). Stripping significant
enough to truncate a massive galaxy to 15-20 kpc may happen even in
the low ICM density outskirts of clusters (Roediger \& Hensler 2004).
The ram pressure model of Vollmer \etal ~(2001), which is based on the
relatively low density ICM case of the Virgo Cluster, predicts that
stripping may occur over more than one orbit of the cluster, depending
on the orbit and the galaxy orientation.  Many of these models also
predict that star formation in the bound gas can be enhanced, either
through a compression directly due to the ICM gas (e.g., Dressler \&
Gunn 1983; Gavazzi \etal ~2001) or by displaced gas falling back into
the disk (e.g., Vollmer \etal ~2001).  The 
models of Vollmer \etal, for
example, predict an enhancement of the star formation rate within the
central 0.25$r_{25}$ of the disk, caused by a passage close to the
cluster center.
These models can account for the many HI deficient galaxies in the
Virgo Cluster, but they may also provide an explanation of why
relatively HI normal galaxies could show a steepening/truncation of
the H$\alpha$ profiles, indicating a modification in their gas content
within the last 10 Myr.  The galaxies in our sample that are HI-normal
but H$\alpha$-small could be those that have passed previously
through the core but were not completely stripped (Vollmer et
al. 2001).  Of the Virgo cluster galaxies in this sample, at least two show
signs of a recent ICM-ISM interaction, including NGC~4522 (Kenney \&
Koopmann 1999; Kenney, van Gorkom, \& Vollmer 2004) and
NGC~4654 (Phookun \& Mundy 1995).  Galaxies that are
located relatively far from the Virgo core, however, are likely more
strongly affected by other environmental processes.

Another process that selectively removes gas is starvation or
strangulation, i.e., the stripping of gas from a galaxy's surroundings
that might otherwise have accreted onto the galaxy (Larson, Tinsley,
\& Caldwell 1980; Balogh, Navarro, \& Morris 2000). This
mechanism could operate far from the cluster core. However, simulations
of Bekki \etal ~(2002) show that starvation leads to gas and star
formation distributions that closely resemble anemic spirals, with
lowered rates of star formation throughout the disk.  It is thus not
clear how starvation/strangulation might produce a shorter H$\alpha$
scale length.

Gas contents and distributions can also be affected by gravitational effects 
(Toomre \& Toomre 1972; reviews by Struck 1999 and Mihos 2004) that
can occur throughout a cluster, including
(i) relatively slow interactions within cluster substructures,
(ii) galaxy harassment due to high-velocity tidal interactions and collisions
(e.g., Moore, Lake, \& Katz 1998),
or (iii) tidal interactions between galaxies and the cluster as a whole
(Byrd \& Valtonen 1990).
Models of galaxy tidal interactions predict significant relocations of
gas, both toward inner and outer radii (e.g., Barnes \& Hernquist
1991).  Lake, Katz and Moore (1998) show that galaxy harassment tends
to drive gas towards the central regions. A resulting increase in star
formation in the central regions due to an enhanced rate of
cloud-cloud collisions would cause a steepening of the H$\alpha$
profile.  In agreement with the models, star formation rates have been
observed to be statistically higher for interacting galaxies (e.g.,
Kennicutt 1998).  Tidal interactions, in fact, appear to be fairly
common in the Virgo Cluster, as shown by a number of kinematic (Rubin
\etal ~1999; Dale \etal ~2001) and imaging (e.g., Koopmann \& Kenney
2004b) studies.

In the absence of detailed modeling of the evolution of the radial
distribution of star formation for the different types of
environmental processes, the above comparisons can be viewed as only
suggestive. We conclude that the smaller H$\alpha$ scale lengths of
Virgo cluster galaxies indicate an additional constraint on the types
of environmental processes that cause gas depletion and a suppression
of the star formation rate in clusters of galaxies.

\acknowledgments
This work was supported by NSF grant AST-0307396. We thank Karen Masters
for use of her flow model program to calculate distances for the field
galaxies. RAK would like to thank the Cornell University Astronomy Department 
for its hospitality to her as a sabbatic visitor.

\clearpage
\begin{landscape}

\begin{deluxetable}{llccrrrrrrcccr}
\tabletypesize{\scriptsize}
\tablecaption{Properties of Sample Galaxies\label{tabprop}}
\tablewidth{0pt}
\tablehead{
\colhead{(1)}& \colhead{(2)}&\colhead{(3)} &\colhead{(4)} &\colhead{(5)}
&\colhead{(6)} & \colhead{(7)} & \colhead{(8)} & \colhead{(9)}&
\colhead{(10)}
& \colhead{(11)}& \colhead{(12)}& \colhead{(13)}& \colhead{(14)}\\
\colhead{NGC/IC}&
\colhead{Other}&
\colhead{RSA}&
\colhead{RC3}&
\colhead{B$_T^O$}&
\colhead{b/a($i$)}&
\colhead{Dist}&
\colhead{$\langle {\rm HI} \rangle$}&
\colhead{r$_{H\alpha95}$}&
\colhead{r$_{R}$}&
\colhead{r$_{\alpha}$(1-3$r_R$)}&
\colhead{r$_{\alpha}$(1-2$r_R$)}&
\colhead{r$_{\alpha}$($r_R-r_s$)}&
\colhead{r$_{max}$}\\
\colhead{}&
\colhead{}&
\colhead{}&
\colhead{}&
\colhead{}&
\colhead{--- ($^{\circ}$)}&
\colhead{(Mpc)}&
\colhead{}&
\colhead{($^{\prime\prime}$)}&
\colhead{($^{\prime\prime}$)}&
\colhead{($^{\prime\prime}$)}&
\colhead{($^{\prime\prime}$)}&
\colhead{($^{\prime\prime}$)}&
\colhead{($^{\prime\prime}$)}\\
\hline\\[-3pt]
\multicolumn{13}{c}{\bf Virgo}
}
\startdata
N 4064 & U 7054 &SBc:  &SBa:pec  &12.30 &0.391(70)&16.6 & 1.23&  25&    44$\pm$2 &    \dots&    \dots  &       \dots&\dots\\
N 4178 & U 7215 &SBc   &SBdm     &11.89 &0.342(74)&16.6 &$-$0.09& 120&  45$\pm$4&96$\pm$20\dag&120$\pm$30\dag&95$\pm$30\dag &97\\
N 4189 & U 7235 &SBc   &SABcd    &12.53 &0.695(47)&16.6 & 0.27&  58&    20$\pm$5&   $\ast$ &  $\ast$ &30$\pm$10\dag &61\\
N 4192 & U 7231 &Sb:   &SABab    &10.92 &0.276(79)&16.6 & 0.18& 194&    72$\pm$4& 58$\pm$ 5&  $\ast$   &72$\pm$10   & 228\\
N 4212 & U 7275 &Sc    &SAc:      &11.86 &0.643(51)&16.6 & 0.16&  55&    22$\pm$4&     \dots&  $\ast$   &15$\pm$4\dag&55\\
N 4237 & U 7315 &Sc    &SABbc    &12.37 &0.682(48)&16.6 & 0.50&  44&    16$\pm$4&     \dots&  $\ast$   &10$\pm$3\dag&40\\
N 4254 & U 7345 &Sc    &SAc       &10.43 &0.839(34)&16.6 &$-$0.14& 123&  33$\pm$3& 41$\pm$ 4&  55$\pm$ 5&33$\pm$3    &\dots\\
N 4293 & U 7405 &Sa pec&SB0/a pec&11.20 &0.438(67)&16.6 & 1.5 &  44&    60$\pm$5&     \dots&      \dots&       \dots&\dots\\
N 4294 & U 7407 &SBc   &SBcd     &12.62 &0.391(70)&16.6 &$-$0.37&  51&  22$\pm$2& 12$\pm$ 2&  15$\pm$ 2&   14$\pm$1 & 64\\
N 4298 & U 7412 &Sc    &SAc       &12.08 &0.530:(60)&16.6& 0.08&  71&    32$\pm$3&     \dots&  25$\pm$ 2&   25$\pm$1 &67\\
N 4299 & U 7414 &Scd   &SABdm:   &12.86 &0.927:(22)&16.6&$-$0.22&  28&  15$\pm$4&     \dots&7$\pm$ 2\dag&    6$\pm$1 &\dots\\
N 4303 & U 7420 &Sc    &SABbc    &10.17 &0.906(26)&16.6 &$-$0.03& 133&  42$\pm$3& 24$\pm$ 4&     $\ast$&   28$\pm$3 &\dots\\
N 4321 & U 7450 &Sc    &SABbc    &10.11 &0.875(30)&16.6 & 0.21& 148&    75$\pm$4& 30$\pm$ 5&  58$\pm$ 5&   41$\pm$5 & 129\\
N 4351 & U 7476 &Sc    &SBab:pec &13.04 &0.695(47)&16.6 & 0.37&  30&    25$\pm$5&     \dots&      \dots&     \dots  &\dots\\
N 4380 & U 7503 &Sab   &SAb:      &12.36 &0.530(60)&16.6 & 0.66&  63&    35$\pm$5&     \dots&     $\ast$&     \dots  &\dots\\
N 4383 & U 7507 &Amorph&SAa pec   &12.68 &0.643(51)&16.6 &$-$0.53&91&       \dots&     \dots&      \dots&       \dots&\dots\\
N 4394 & U 7523 &SBb   &RSBb?    &11.76 &0.906(26)&16.6 & 0.39&  89&    33$\pm$2&     \dots&   $\ast$  & 50$\pm$10  &\dots\\
N 4405 & U 7529 &Sc/S0 &SA0/a    &12.99 &0.695(47)&16.6 & 0.85&  19&    18$\pm$5&     \dots&      \dots&   \dots    &\dots\\
N 4413 & U 7538 &SBbc  &SBab:    &12.97 &0.656(50)&16.6 & 0.26&  34&    20$\pm$2&     \dots&      \dots& \dots &  34\\
N 4411B& U 7546 &Sc    &SABcd    &12.92 &1.00(0)&16.6   & 0.60&  76&    29$\pm$2&     \dots& 30$\pm$ 4& 32$\pm$3   &\dots\\
N 4419 & U 7551 &Sa    &SBa      &12.13 &0.391(70)&16.6 & 0.78&  46&    20$\pm$2&     \dots& 19$\pm$ 2& 18$\pm$2   &  44\\
N 4424 & U 7561 &Sa pec&SAB0p    &12.32 &0.616(54)&16.6 & 0.79&  27&    33$\pm$4&     \dots&      \dots&       \dots&\dots\\
N 4450 & U 7594 &Sab pec&SAab    &10.93 &0.707(46)&16.6 & 0.99&  92&    41$\pm$4&     \dots&      \dots&       \dots&\dots\\
I 3392 & U 7602 &Sc/Sa &SAb:     &13.30 &0.454(65)&16.6 & 0.96&  30&    22$\pm$3&     \dots&      \dots&       \dots&\dots\\ 
N 4457 & U 7609 &RSb   &SAB0/a   &11.76 &1.00:(0)&16.6  & 0.92&  33&    40$\pm$3&     \dots&      \dots&       \dots&\dots\\
N 4498 & U 7669 &SBc   &SABd     &12.62 &0.485(63)&16.6 & 0.16&  70&    25$\pm$2& 19$\pm$ 2&  23$\pm$ 2& 20$\pm$3   &  63\\
N 4501 & U 7675 &Sbc   &SAb      &10.27 &0.50(62)&16.6  & 0.34& 112&    46$\pm$3& 32$\pm$ 4&  42$\pm$ 4& 40$\pm$2   & 105\\
N 4519 & U 7709 &SBc   &SBd      &12.34 &0.743(43)&16.6 &$-$0.18&  86&  21$\pm$4&    $\ast$&    $\ast$ &30$\pm$10\dag& 157\\
N 4522 & U 7711 &Sc/Sb:&SBcd:    &12.73 &0.276(79)&16.6 & 0.69&  54&    32$\pm$3& 14$\pm$ 2& 8$\pm$ 2\dag&  11$\pm$1&  43\\
N 4527 & U 7721 &Sb    &SABbc    &11.32 &0.309(76)&16.6 &$-$0.19& 138&  50$\pm$5& 38$\pm$ 2&  60$\pm$10& 38$\pm$2   & 160\\
N 4532 & U 7726 &Sm    &Im       &12.30 &0.391(70)&16.6 &$-$0.35&  68&  18$\pm$3& 15$\pm$ 1&  14$\pm$ 2& 13$\pm$1   & 40\\
N 4535 & U 7727 &SBc   &SABc     &10.51 &0.731(44)&16.6 & 0.16& 156&    60$\pm$6& 50$\pm$10&   $\ast$  &77$\pm$10   &148\\
N 4536 & U 7732 &Sc    &SABbc    &11.01 &0.391(70)&16.6 & 0.17& 198&    68$\pm$7& 67$\pm$ 10&  66$\pm$ 3& 67$\pm$10 & 220\\
N 4548 & U 7753 &SBb   &SBb      &10.98 &0.809(37)&16.6 & 0.76& 126&    55$\pm$5&     \dots&  70$\pm$20\dag&60$\pm$20\dag&  109\\
N 4561 & U 7768 &SBc   &SBdm     &12.96:&0.829(35)&16.6 &$-$0.64&  34&  18$\pm$4&     \dots&   8$\pm$ 2\dag&8$\pm$2\dag&\dots\\
N 4567 & U 7777 &Sc    &SAbc     &12.08 &0.695(47)&16.6 & 0.64&  41&    31$\pm$3&     \dots&      \dots& \dots      &  43\\
N 4568 & U 7776 &Sc    &SAbc     &11.70 &0.438(67)&16.6 & 0.64&  72&    36$\pm$3&     \dots&  28$\pm$ 5& 18$\pm$2   &   86\\
N 4569 & U 7786 &Sab   &SABab    &10.25 &0.469(64)&16.6 & 0.92&  65&    66$\pm$4&     \dots&      \dots&       \dots&\dots\\
N 4571 & U 7788 &Sc    &SAd      &11.81 &0.829(35)&16.6 & 0.49&  92&    40$\pm$4&     \dots&  70$\pm$20\dag&40$\pm$4&    87\\
N 4579 & U 7796 &Sab   &SABb     &10.56 &0.799(38)&16.6 & 0.63&  90&    42$\pm$3&     \dots&  46$\pm$ 5    &46$\pm$ 5& 96\\
N 4580 & U 7794 &Sc/Sa &SABa?    &12.49 &0.719(45)&16.6 & 1.3 &  23&    19$\pm$5&     \dots&      \dots&       \dots&\dots\\
N 4606 & U 7839 &Sa pec&SBa:     &12.69 &0.438(67)&16.6 & 1.2 &  47&    25$\pm$5&     \dots&      \dots&       \dots&\dots\\
N 4639 & U 7884 &SBb   &SABbc    &12.19 &0.669(49)&16.6 & 0.18&  67&    20$\pm$2& 20$\pm$ 3&  $\ast$   &   27$\pm$4 &   81\\
N 4647 & U 7896 &Sc    &SABc     &12.03 &0.809(37)&16.6 & 0.43&  57&    25$\pm$3& 13$\pm$ 2&  25$\pm$ 5& 13$\pm$2   &   49\\
N 4651 & U 7901 &Sc    &SAc      &11.36 &0.642(51)&16.6 &$-$0.28&  92&  35$\pm$5& 30$\pm$ 4&  30$\pm$ 4& 30$\pm$4   &  101\\
N 4654 & U 7902 &SBc   &SABcd    &11.14 &0.629(52)&16.6 & 0.05& 101&    36$\pm$3& 23$\pm$ 2&  24$\pm$ 4& 27$\pm$3   &  97\\
N 4689 & U 7965 &Sc    &SAbc     &11.55 &0.819(36)&16.6 & 0.57&  64&    50$\pm$5&     \dots&      \dots&       \dots&  51\\
N 4694 & U 7969 &Amorph&SB0 pec  &12.19 &0.755(42)&16.6 & 0.81&  25&    40$\pm$3&     \dots&      \dots&       \dots&\dots\\
N 4698 & U 7970 &Sa    &SAab     &11.53 &0.500(62)&16.6 &$-$0.14& 247&  70$\pm$5&     \dots&      \dots&       \dots&89\\
N 4713 & U 7985 &SBc   &SABd     &12.21 &0.656(50)&16.6 &$-$0.35&  63&  17$\pm$4& 25$\pm$ 5&  $\ast$   & 18$\pm$2    & 35\\
N 4772 & U 8021 &...   &SAa      &11.89 &0.500(62)&16.6 & 0.07&  72&    42$\pm$4&     \dots&      \dots& \dots   &\dots\\
N 4808 & U 8054 &Sc    &SAbc:    &12.56:&0.391(70)&16.6 &$-$0.68&  65&  19$\pm$4& 16$\pm$ 1&  35$\pm$10\dag& 15$\pm$2&   71\\
\enddata
\tablecomments{\dag Scale lengths with measurement errors larger than 20\%.
$\ast$ Scale length fit produced a negative or zero value (clearly
unphysical). }
\end{deluxetable}
\clearpage

\setcounter{table}{0}
\begin{deluxetable}{llccrrrrrrcccr}
\tabletypesize{\scriptsize}
\tablecaption{Continued}
\tablewidth{0pt}
\tablehead{
\colhead{(1)}& \colhead{(2)}&\colhead{(3)} &\colhead{(4)} &\colhead{(5)}
&\colhead{(6)} & \colhead{(7)} & \colhead{(8)} & \colhead{(9)}&
\colhead{(10)}
& \colhead{(11)}& \colhead{(12)}& \colhead{(13)}& \colhead{(14)}\\
\colhead{NGC/IC}&
\colhead{Other}&
\colhead{RSA}&
\colhead{RC3}&
\colhead{B$_T^O$}&
\colhead{b/a($i$)}&
\colhead{Dist}&
\colhead{$\langle {\rm HI} \rangle$}&
\colhead{r$_{H\alpha95}$}&
\colhead{r$_{R}$}&
\colhead{r$_{\alpha}$(1-3r$_R$)}&
\colhead{r$_{\alpha}$(1-2r$_R$)}&
\colhead{r$_{\alpha}$($>$r$_R$)}&
\colhead{r$_{max}$}\\
\colhead{}&
\colhead{}&
\colhead{}&
\colhead{}&
\colhead{}&
\colhead{--- ($^{\circ}$)}&
\colhead{(Mpc)}&
\colhead{}&
\colhead{($^{\prime\prime}$)}&
\colhead{($^{\prime\prime}$)}&
\colhead{($^{\prime\prime}$)}&
\colhead{($^{\prime\prime}$)}&
\colhead{($^{\prime\prime}$)}&
\colhead{($^{\prime\prime}$)}\\
\hline\\[-3pt]
\multicolumn{13}{c}{\bf Field}
}
N  578 & UA 18  &Sc   &SABc     &11.17 &0.574(57)&26.2 &$-$0.14&  104& 35$\pm$3& 42$\pm$ 3&  50$\pm$ 5 & 27$\pm$ 2  &\dots\\
N  613 & 413G11 &SBb  &SBbc     &10.53 &0.799(38)&23.9 & 0.06&  101&   63$\pm$3&     \dots&    $\ast$  & 30$\pm$5   &\dots\\
N  628 & U 1149 &Sc   &SAc      & 9.76 &0.857(32)&7.3  &$-$0.11&  242& 72$\pm$3&100$\pm$10& 110$\pm$10& 80$\pm$ 4   &\dots\\
N  925 & U 1913 &SBc  &SABd     & 9.97 &0.629(52)&9.1  & 0.01&  220&   66$\pm$5& 90$\pm$ 5& 130$\pm$10&120$\pm$20   &\dots\\
N  986 & 299G7  &SBb  &SBab      &11.45 &0.743(43)&32.2 & 0.40&  82&    38$\pm$3&     \dots&      \dots&50$\pm$20\dag&\dots\\
N 1087 & U 2245 &Sc   &SABc     &10.97 &0.616(53)&23.5 & 0.04&   60&   31$\pm$2&     \dots&      \dots&20$\pm$4     &  73\\
N 1169 & U 2503 &SBa  &SABb     &10.86 &0.682(48)&34.4 &$-$0.25&  103& 32$\pm$3&  $\ast$  &     $\ast$&120$\pm$50\dag& 74\\
N 1232 & 547G14 &Sc  &SABc     &10.38 &0.848(33)&25.8 &$-$0.07&  182& 55$\pm$5& 75$\pm$10& 120$\pm$10& 55$\pm$5    & 165\\
N 1249 & 155G6  &SBc  &SBcd     &11.64 &0.500(62)&17.4 &$-$0.05&  137& 40$\pm$3& 21$\pm$ 1&  29$\pm$ 2& 30$\pm$5    &  148\\
I  356 & U 2953 &\dots&SAab pec &10.17 &0.719(45)&14.6 &$-$0.47&  184& 81$\pm$10& 40$\pm$10\dag&82$\pm$5&50$\pm$10  &\dots\\
N 1637 & UA 93  &SBc  &SABc     &11.25 &0.755(42)&11.7 &$-$0.24&   96& 30$\pm$4& 31$\pm$ 4&  40$\pm$ 5&28$\pm$1     &\dots\\
N 1832 & M-0314010 &SBb &SBbc   &11.59 &0.669(49)&30.8 &$-$0.32&   39& 17$\pm$1& 60$\pm$20\dag&  60$\pm$40\dag&18$\pm$4\dag& 77\\
N 2090 & 363G23 &Sc   &SAc   &11.45 &0.484(63)&15.7 & 0.07&  203&        \dots&     \dots&    \dots  &  \dots  &96\\
N 2196 & UA 121 &Sab  &SAab    &11.38 &0.682(48)&36.8 &$-$0.44&  112& 31$\pm$5& 46$\pm$2 & 100$\pm$30\dag&41$\pm$2 &  74\\
 \dots & U 3580 &\dots&SAa pec: &12.20 &0.469(64)&18.7 &$-$0.24&  100& 28$\pm$3& 24$\pm$5\dag&  11$\pm$ 2&31$\pm$3  &  39\\
N 2403 & U 3918 &Sc   &SABcd    & 8.43 &0.559(58)&3.2  &\dots&  544&  110$\pm$4&100$\pm$10 &   $\ast$     &110$\pm$10&\dots\\
N 2525 & UA 135 &SBc  &SBc      &11.55 &0.643(51)&25.5 & 0.08&   69&   30$\pm$5&     \dots&  $\ast$   &15$\pm$4\dag &\dots\\
N 2608 & U 4484 &Sbc  &SBb      &12.53 &0.573(57)&31.2 & 0.36&  38 &   20$\pm$2&     \dots&         \dots& \dots  &\dots\\
N 2712 & U 4708 &SBb  &SBb:     &12.19 &0.500(62)&26.4 & 0.01&   69&   19$\pm$2&    $\ast$&     $\ast$&20$\pm$10\dag&   82\\
N 2715 & U 4759 &Sc   &SABc     &11.09 &0.342(74)&21.0 & 0.08&  123&   35$\pm$3& 72$\pm$10&  43$\pm$ 5&60$\pm$8     &  124\\
N 2805 & U 4936 &\dots&SABd     &11.17 &0.755(42)&25.6 & 0.05&  181&   60$\pm$5& 72$\pm$ 4& 100$\pm$20   &85$\pm$10 &\dots\\
N 2841 & U 4966 &Sb   &SAb      & 9.58 &0.454(65)&10.8 &$-$0.13&  251& 58$\pm$4& 62$\pm$ 5& 140$\pm$30\dag&62$\pm$5&  179\\
N 2903 & U 5079 &Sc   &SBd      & 9.11 &0.515(61)&8.9  & 0.07&  147&   68$\pm$6& 31$\pm$ 4&  40$\pm$ 5   &35$\pm$3  &\dots\\
N 3031 & U 5318 &Sb   &SAab     & 7.39 &0.515(61)&3.6  & 0.99&  565&  160$\pm$10&600$\pm$200\dag&  $\ast$&300$\pm$100\dag& 140\\
N 3198 & U 5572 &Sc   &SBc      &10.21 &0.358(72)&13.7 &$-$0.14&  247& 50$\pm$5& 70$\pm$10&  95$\pm$15& 65$\pm$ 5   & 196\\
N 3329 & U 5837 &Sab &SAb:     &12.57 &0.559(58)&27.2 &\dots&   53&   20$\pm$3& 15$\pm$ 1&   8$\pm$ 2\dag&15$\pm$1 &  62\\
N 3359 & U 5873 &SBc pec&SBc    &10.68 &0.643(51)&16.0 &$-$0.22&  176& 43$\pm$2& 46$\pm$ 2&  43$\pm$ 3 & 45$\pm$1   & 213\\
N 3627 & U 6346 &Sb   &SABb     & 9.13 &0.500(62)&9.4  & 0.46&  145&   70$\pm$10&     \dots&  33$\pm$ 2&31$\pm$2    &\dots\\
N 3673 & UA 236 &SBb  &SBb      &11.81 &0.500(62)&31.6 & 0.04&  124&   33$\pm$3&36$\pm$20\dag&  $\ast$ &42$\pm$4    & 76\\
N 3705 & U 6498 &Sb   &SABab    &11.25 &0.423(68)&16.5 & 0.67&  141&   32$\pm$2& 31$\pm$ 5&  30$\pm$ 5&    32$\pm$ 1& 83\\
N 3887 & UA 246 &SBbc &SBbc     &11.19 &0.766(41)&20.2 &$-$0.21&  100& 27$\pm$3&    $\ast$&     $\ast$& 25$\pm$ 4   & 118\\
N 4395 & U 7524 &Sd   &SAm      &10.57 &0.777(40)&4.0  & 0.13&  178&  130$\pm$20&     \dots&     \dots& \dots       &\dots\\
N 4597 & M-0132034 &SBc: &SBm   &12.21 &0.423(68)&17.4 &$-$0.23&  123& 40$\pm$6&     $\ast$&    $\ast$& 120$\pm$60\dag& 109\\
N 4800 & U 8035 &Sb   &SAb      &12.13 &0.766(41)&20.2 & 0.21&   40&   19$\pm$5& 10$\pm$ 1&  10$\pm$ 2&  10$\pm$1   & 39\\
N 4984 & M-0234004 &Sa &SAB0+   &12.03 &0.755(42)&21.3 & 0.76&   35&   40$\pm$4&     \dots&      \dots&        \dots&\dots\\
N 5248 & U 8616 &Sbc  &SBbc     &10.63 &0.574(57)&18.5 &$-$0.11&  265& 40$\pm$8& 30$\pm$ 3& $\ast$    &35$\pm$5     &\dots\\
N 5334 & U 8790 &SBc  &SBc:     &11.62 &0.719(45)&22.2 & 0.11&  115&   40$\pm$4& 45$\pm$ 1&  45$\pm$ 2& 44$\pm$ 1   & 98\\
N 5377 & U 8863 &SBa  &SBa      &11.94 &0.530(60)&25.5 & 0.32&  110&   35$\pm$5& 28$\pm$ 3&    $\ast$& 37$\pm$5    & 55\\
N 5448 & U 8969 &Sa   &SABa     &11.72 &0.423(68)&28.8 &$-$0.16& 126&  26$\pm$3& 43$\pm$ 5&     $\ast$& 37$\pm$2    &  52\\
N 5457 & U 8981 &Sc   &SABcd    & 8.21 &0.934(21)&6.7  &\dots&  744&  140$\pm$10&160$\pm$20& 160$\pm$20&150$\pm$20  &\dots\\
N 5669 & U 9353 &SBc  &SABcd    &11.79 &0.707(46)&21.6 &$-$0.27&  115& 31$\pm$4& 31$\pm$ 5&  45$\pm$ 5&36$\pm$ 3    &  75 \\
N 6118 & U 10350 &Sc  &SAcd     &11.26 &0.438(67)&25.3 & 0.10&  106&   40$\pm$5& 35$\pm$ 5&  55$\pm$10&35$\pm$4     &\dots\\
N 6181 & U 10439 &Sc  &SABc     &11.77 &0.454(65)&35.2 &$-$0.33&   57& 14$\pm$1& 16$\pm$ 2&  $\ast$   &14$\pm$2     & 57\\
N 6643 & U 11218 &Sc  &SAc      &11.14 &0.485(63)&22.8 &$-$0.06&   79& 30$\pm$3& 22$\pm$ 4&  22$\pm$ 4&22$\pm$4     & 74\\
N 7098 &  48G5  &\dots&SABa     &12.23 &0.574(57)&37.5 & 0.83&  126&   42$\pm$4& 72$\pm$ 5&  $\ast$   &72$\pm$5     &\dots\\
N 7141 & 189G7  &\dots&SABbc      &12.21 &0.707(46)&47.7 &\dots&  114&   32$\pm$4&    $\ast$&     $\ast$&50$\pm$10    &\dots\\
N 7177 & U 11872 &Sb  &SABb     &11.47 &0.682(48)&18.1 &$-$0.04&   98& 17$\pm$3& 30$\pm$ 8\dag& $\ast$&27$\pm$2 &  56\\
N 7217 & U 11914 &Sb  &SAab     &10.53 &0.883(29)&15.0 & 0.54&   84&   38$\pm$4&     \dots&     $\ast$&  45$\pm$6   &  52\\
I 5240 & 290G2  &SBa  &SBa     &12.29 &0.643(51)&29.8 & 0.13&   79&   24$\pm$3&    $\ast$&     $\ast$&70$\pm$20\dag&\dots\\
I 5273 & 346G22 &SBc  &SBcd:    &11.55 &0.602(55)&22.2 & 0.08&   60&   31$\pm$2&     \dots&  23$\pm$ 4&23$\pm$4     &\dots\\
N 7448 & U 12294 &Sc  &SAbc     &11.50 &0.454(65)&32.8 &$-$0.38&   66& 16$\pm$2& 28$\pm$ 5&  25$\pm$ 5&13$\pm$5\dag     &  57\\
\enddata
\tablecomments{\dag Scale lengths with measurement errors larger than 20\%.
$\ast$ Scale length fit produced a negative or zero value (clearly
unphysical). }
\end{deluxetable}

\clearpage
\end{landscape}

\appendix
\section{Scale Length Fits}

\setcounter{figure}{0}

\begin{figure*}[h]
\centerline{
\includegraphics{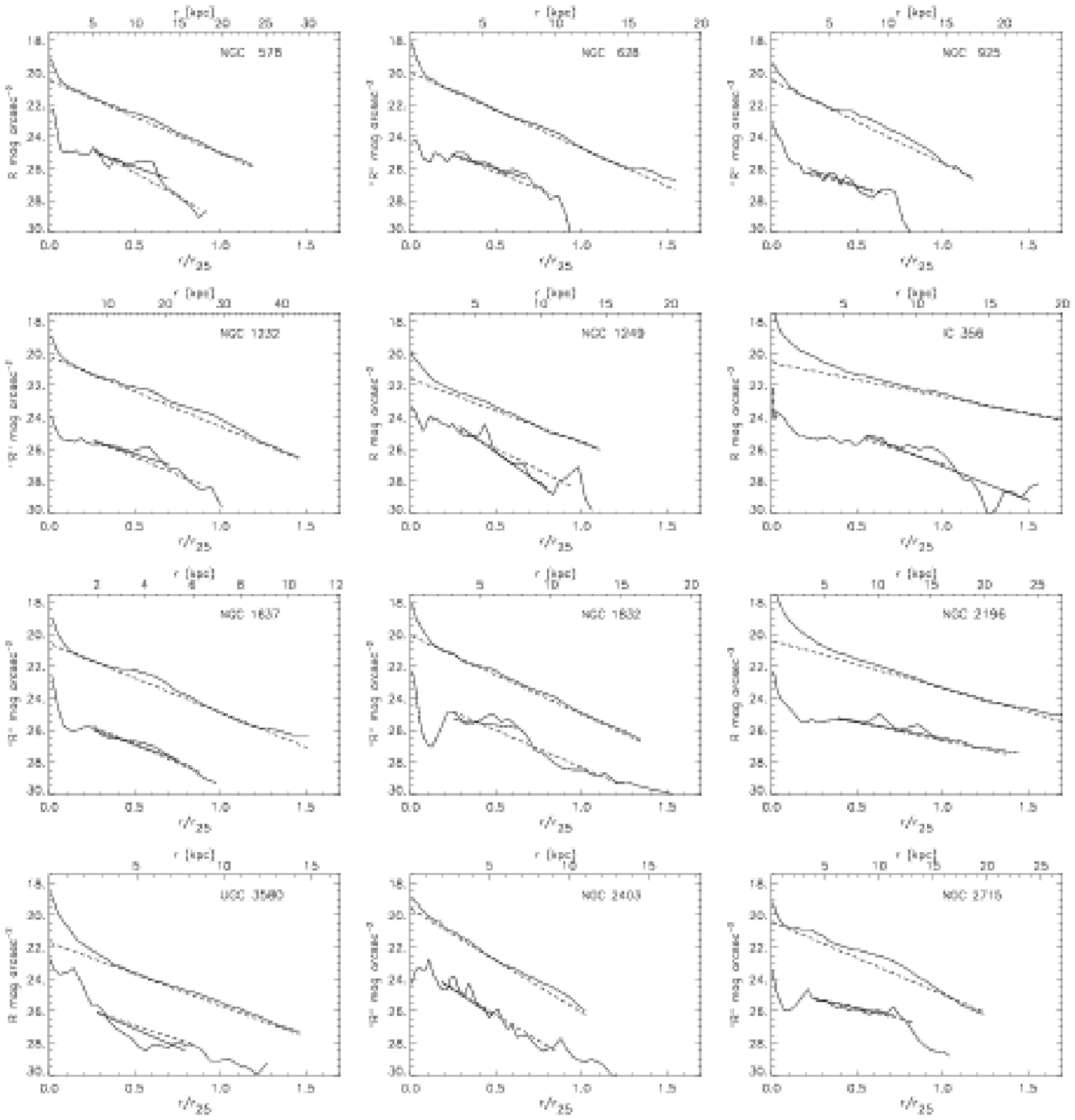}
}
\caption
{R-band and H$\alpha$ surface brightness profiles for the field sample
galaxies in which the H$\alpha$ emission could be traced to r$\geq 3~r_R$.
The H$\alpha$ profiles are plotted below the R-band ones with an arbitrary 
offset. The R-band scale is also arbitrary for galaxies which are not 
calibrated; these galaxies are identified by the ``R'' designation in the 
y-axis. The R-band profiles are plotted to the radius at which the sky
background exceeded the signal.
H$\alpha$ profiles extend to the radius of the last resolved HII region.
The median uncertainty in the H$\alpha$ surface brightness due to the
uncertainty in the sky is 0.3 mag/arcsec$^2$ at a radius of 3$r_R$.
Exponential fits to the R-band and H$\alpha$ profiles are overplotted with
approximate central surface brightnesses. 
H$\alpha$ scale lengths are obtained from fits performed over the
regions indicated in each panel: 1-3$r_R$ (solid) and $r_R-r_s$ (dashed).
\label{scalefig}}
\end{figure*}

\addtocounter{figure}{-1}

\begin{figure*}
\centerline{
\includegraphics{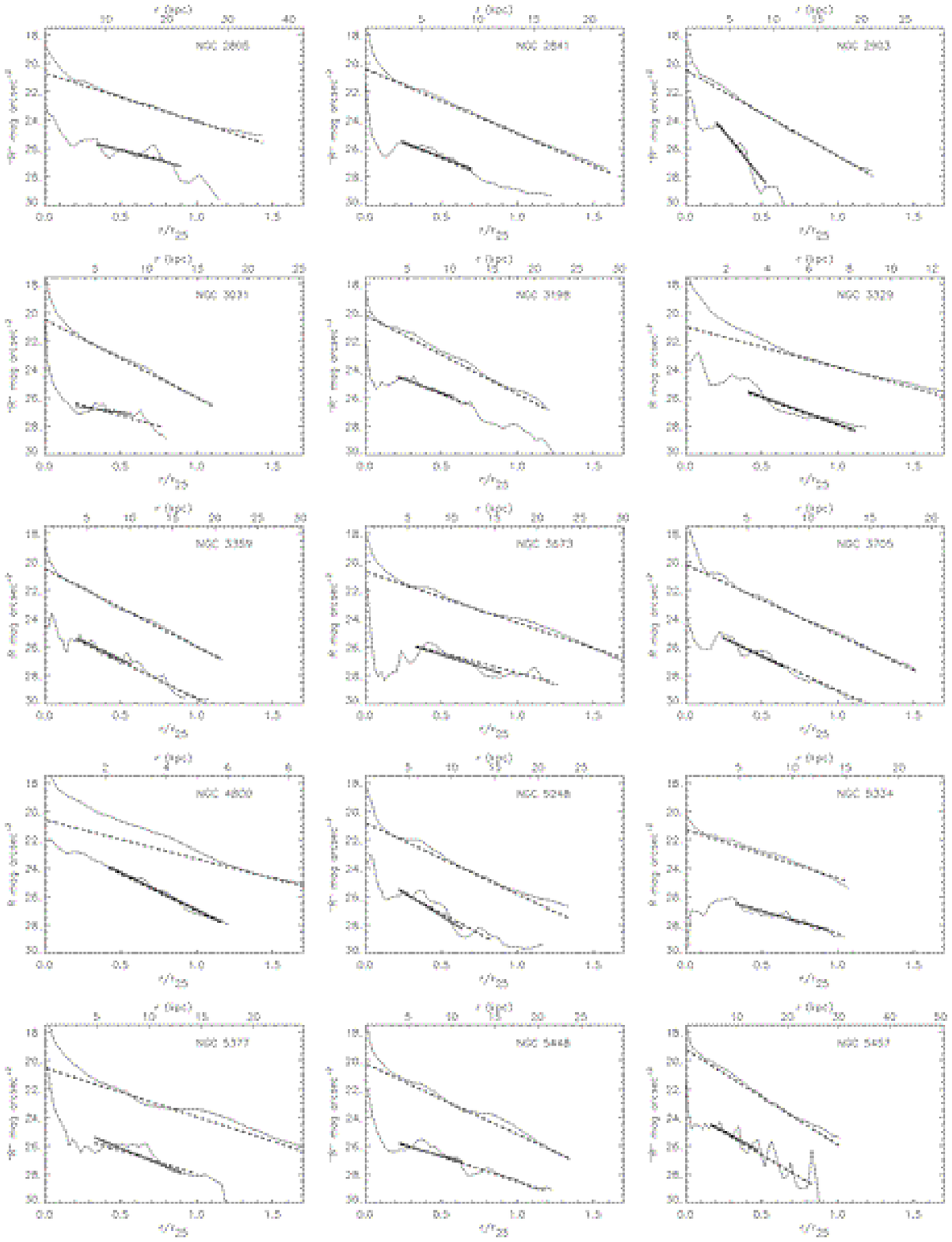}
} 
\caption
{Continued }
\end{figure*}

\clearpage

\addtocounter{figure}{-1}
\begin{figure*}
\centerline{
\includegraphics{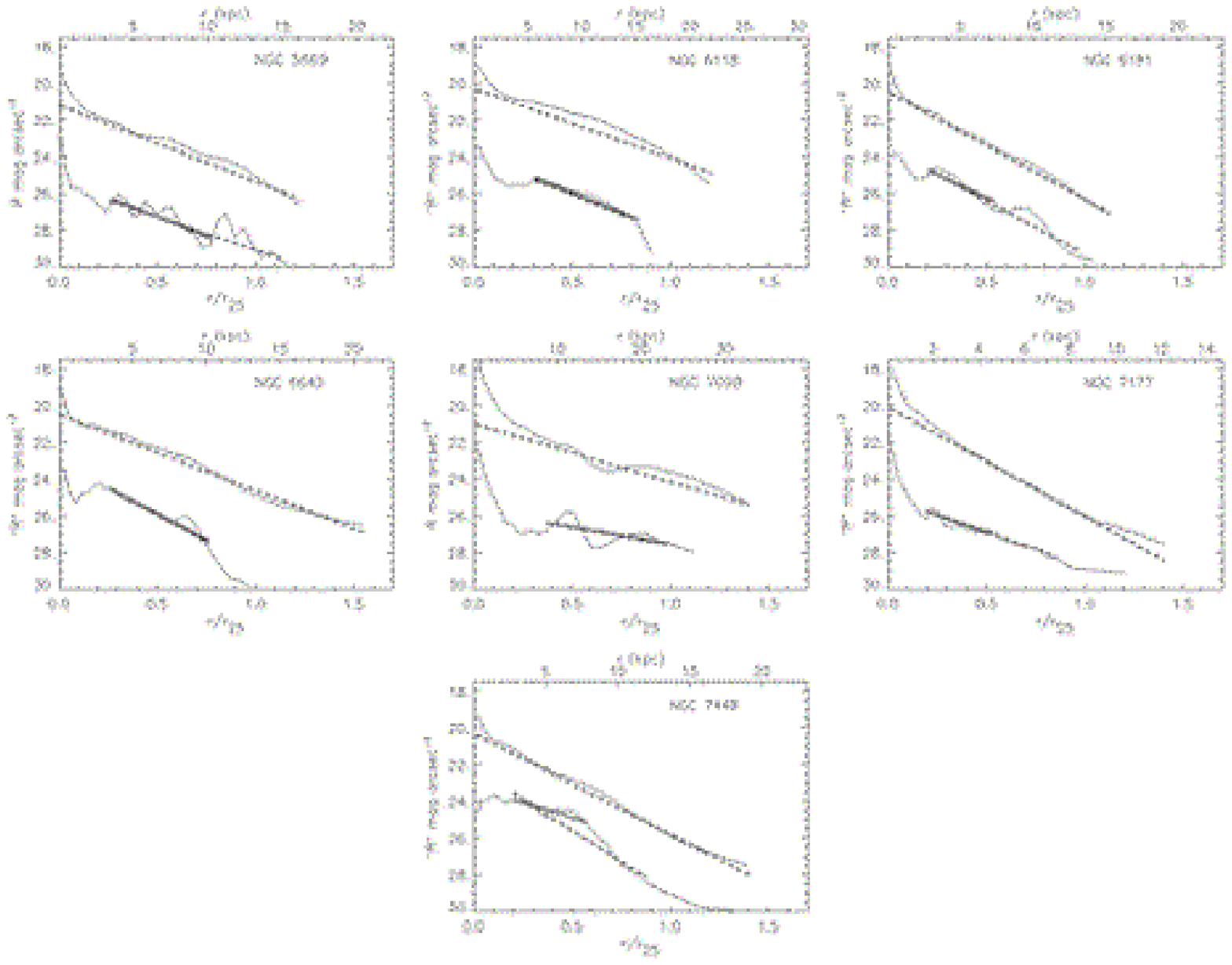}
} 
\caption
{Continued }
\end{figure*}

\clearpage

\begin{figure*}
\centerline{
\includegraphics{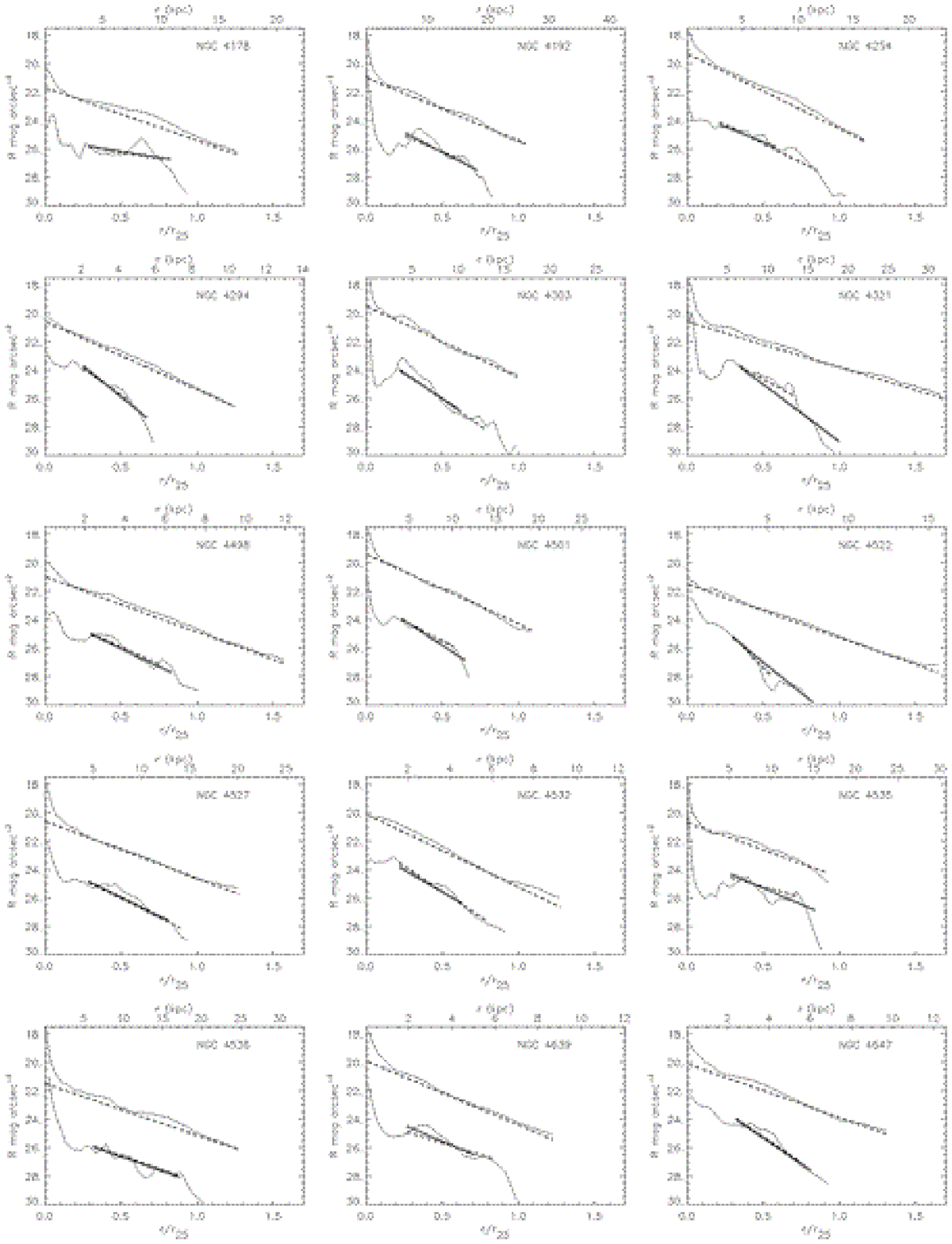}
}
\caption{Same as Figure~\ref{scalefig} for the Virgo Cluster galaxies.
\label{scalefigv}}
\end{figure*}

\clearpage

\addtocounter{figure}{-1}

\begin{figure*}
\centerline{
 \includegraphics{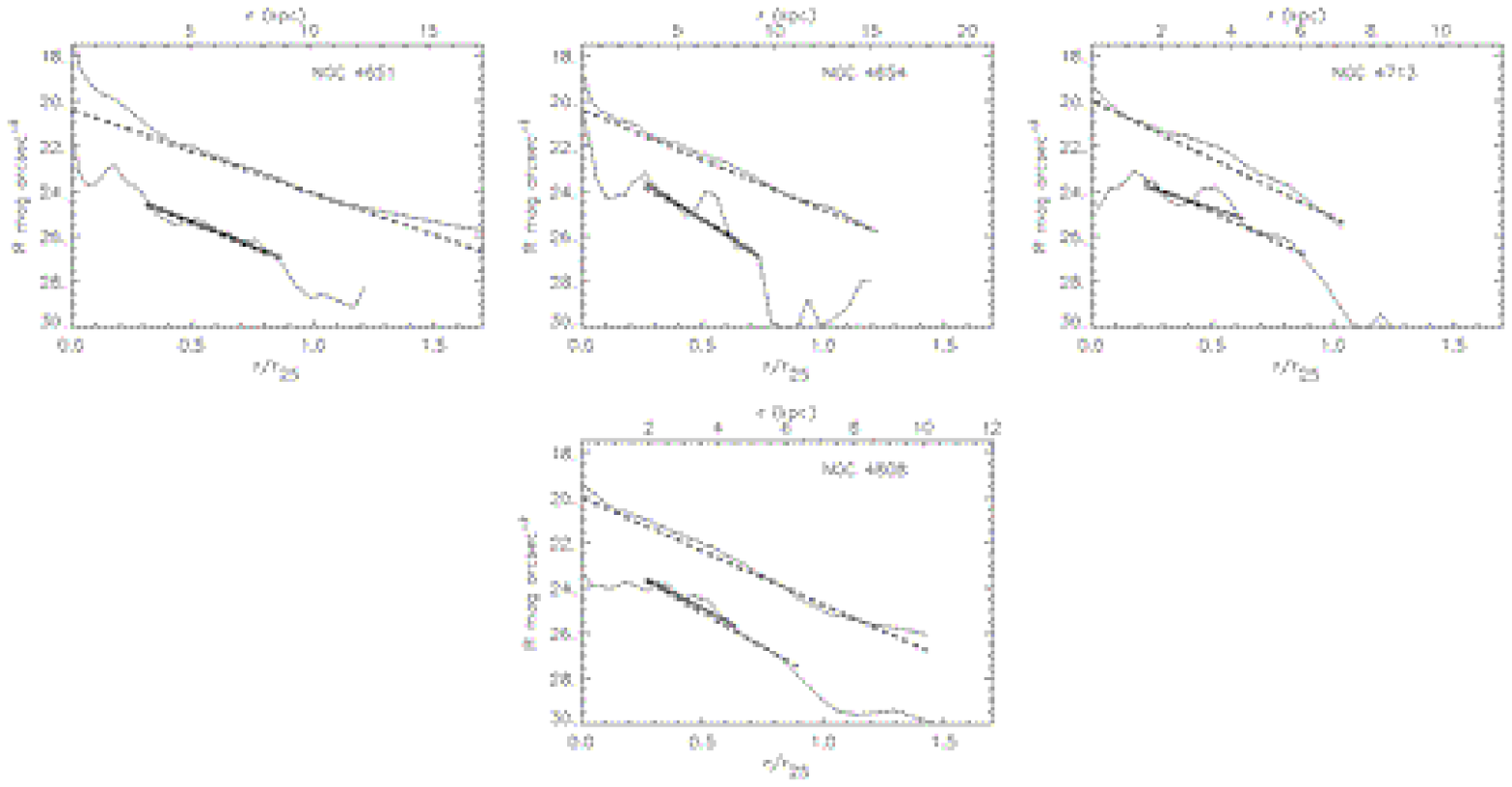}
}
\caption
{Continued }
\end{figure*}

\begin{figure*}
\centerline{
\includegraphics{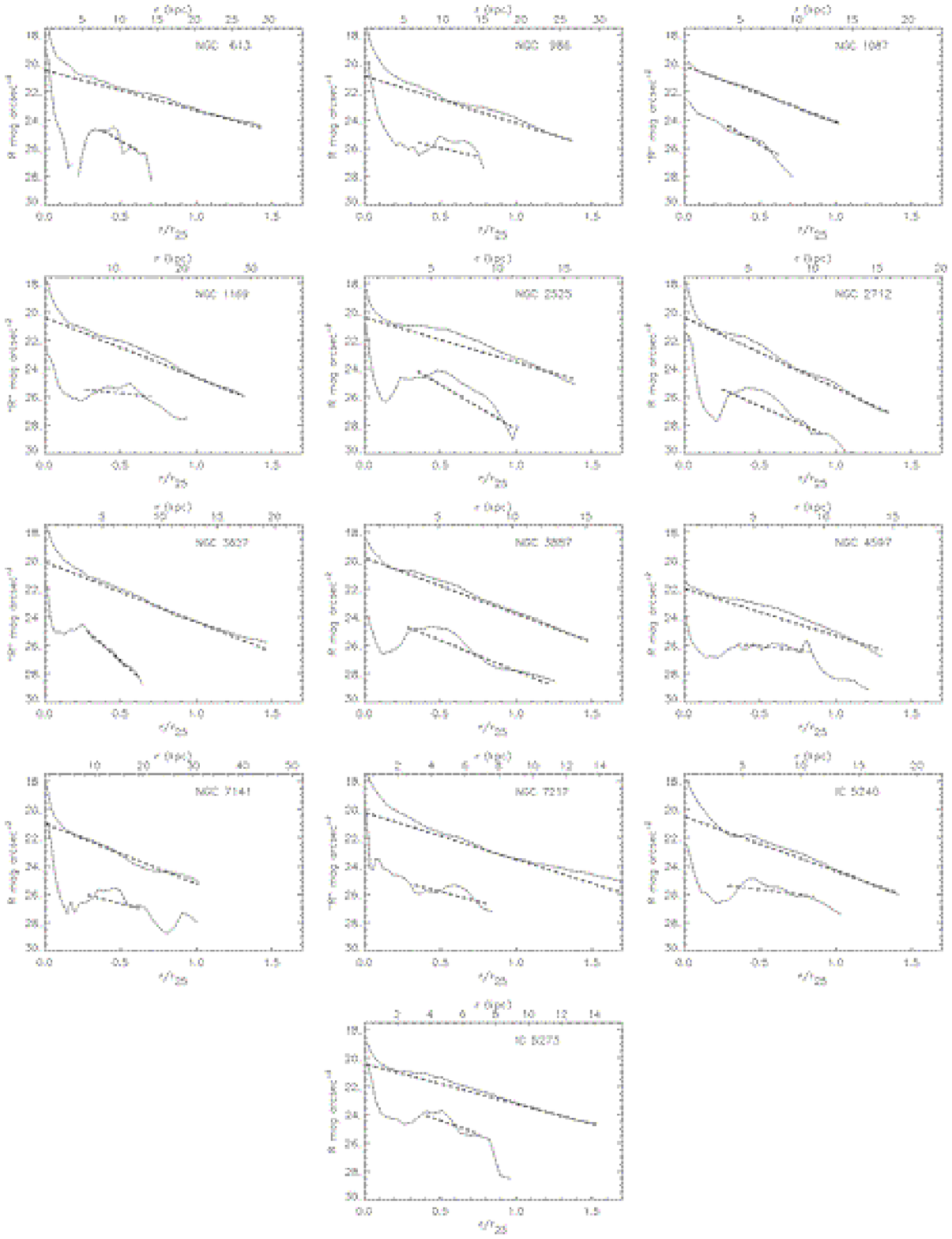}
}
\caption
{R-band and H$\alpha$ surface brightness profiles for 8 field
galaxies in which the H$\alpha$ emission could {\em not} be traced to $3~r_R$.
H$\alpha$ scale lengths for these objects are obtained from
exponential fits to the profiles over the intervals 1-2$r_R$ (not shown)
and $r_R-r_s$ (dashed); in a few cases (NGC 986, 1087, and 2608) only
the latter measurement is available.
See Figure~\ref{scalefig} for plot details.
\label{scalefigtr}}
\end{figure*}

\begin{figure*}
\centerline{
\includegraphics{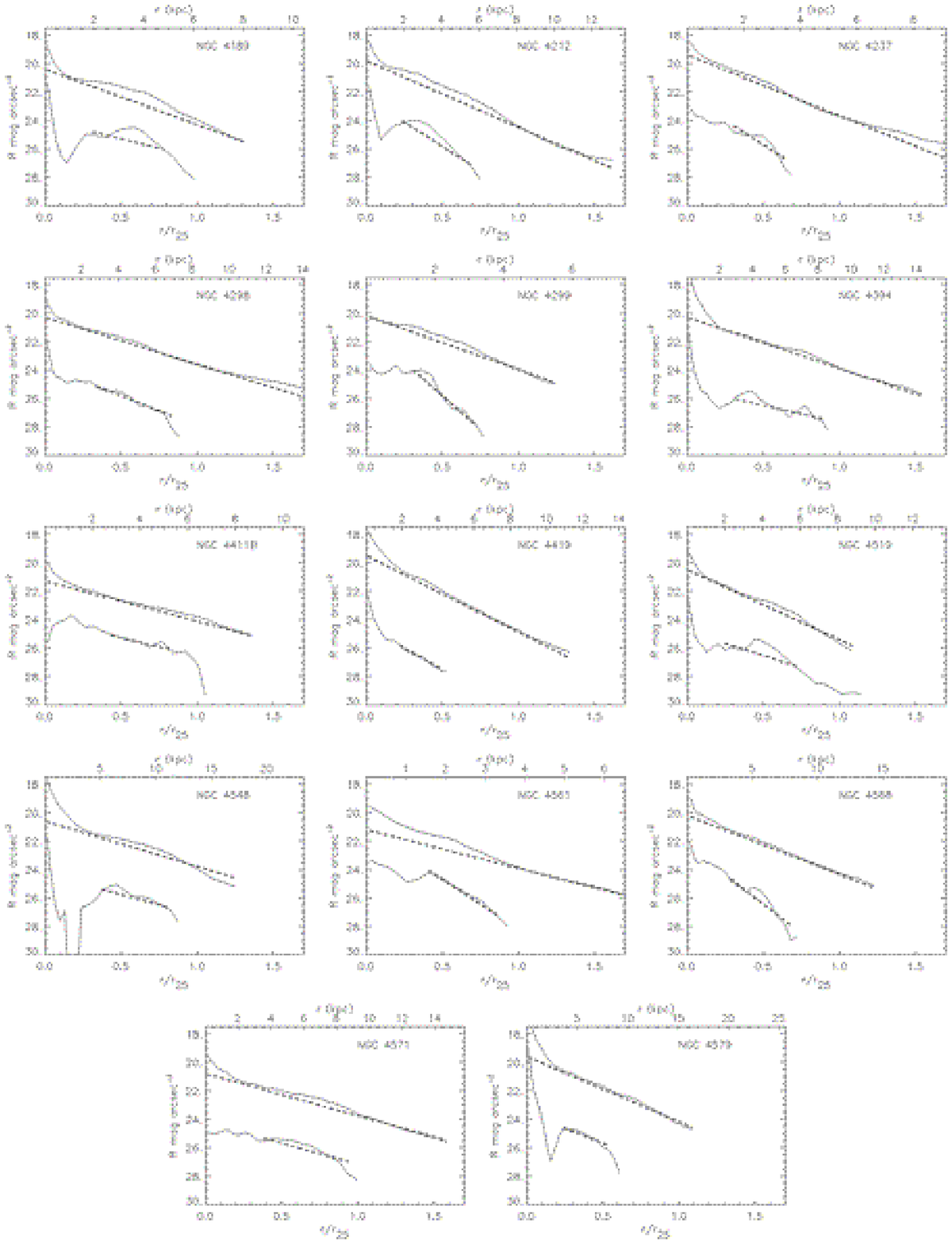}
}
\caption
{Same as Figure~\ref{scalefigtr} for the Virgo Cluster galaxies.
\label{scalefigtrv}}
\end{figure*}

\end{document}